\documentclass[11pt]{elsarticle}
\biboptions{sort&compress,numbers}
\usepackage{microtype}
\usepackage[margin=1in]{geometry}
\usepackage{amsmath,amssymb,amsthm,bm,stmaryrd}
\usepackage{subcaption}
\biboptions{sort&compress}
\usepackage{upgreek}
\usepackage{epstopdf}
\usepackage{float}
\usepackage{xcolor}
\usepackage[labelfont=bf]{caption}
\usepackage[export]{adjustbox}
\AppendGraphicsExtensions{.tif}

\usepackage[final]{changes} 
\definecolor{C0}{HTML}{1F77B4}
\definecolor{C1}{HTML}{FF7F0E}
\definecolor{C2}{HTML}{2ca02c}
\definecolor{C3}{HTML}{d62728}
\definecolor{C4}{HTML}{9467bd}
\definecolor{C5}{HTML}{8c564b}
\definechangesauthor[color=C0]{R1} 
\definechangesauthor[color=C1]{R2} 
\definechangesauthor[color=C3]{RA} 

\begin{document}


    \begin{frontmatter}

        \title{Moving window techniques to model shock wave propagation using the concurrent atomistic-continuum method}
        \author[auburn]{Alexander S. Davis}
        \author[arl]{Jeffrey T. Lloyd}
        \author[auburn]{Vinamra Agrawal\corref{cor1}}
        \ead{vinagr@auburn.edu}
        \cortext[cor1]{Corresponding author}
        \address[auburn]{Department of Aerospace Engineering, Auburn University, Auburn, AL USA}
        \address[arl]{DEVCOM Army Research Laboratory, Aberdeen Proving Ground, MD USA}
        
        \begin{abstract}
            Atomistic methods have successfully modeled different aspects of shock wave propagation in materials over the past several decades, but they suffer from limitations which restrict the total runtime and system size. 
            Multiscale methods have been able to increase the length and time scales that can be modeled but employing such schemes to simulate wave propagation and evolution through engineering-scale domains is an active area of research. 
            In this work, we develop two distinct moving window approaches within a Concurrent Atomistic-Continuum (CAC) framework to model shock wave propagation through a one-dimensional monatomic chain. 
            In the first method, the entire CAC system travels with the shock in a conveyor fashion and maintains the shock front in the middle of the overall domain.
            In the second method, the atomistic region follows the shock by the simultaneous coarsening and refinement of the continuum regions. 
            The CAC and moving window frameworks are verified through dispersion relation studies and phonon wave packet tests. 
            We achieve good agreement between the simulated shock velocities and the values obtained from theory with the conveyor technique, while the coarsen-refine technique allows us to follow the propagating wave front through a large-scale domain. 
            This work showcases the ability of the CAC method to accurately simulate propagating shocks and also demonstrates how a moving window technique can be used in a multiscale framework to study highly nonlinear, transient phenomena.
        \end{abstract}
    
    \end{frontmatter}
    
    \section{Introduction} \label{Sec: Introduction}
        The behavior of shock waves in solids has been extensively studied for many decades \cite{meyers1994dynamic,davison2008fundamentals}.
        Typically, a propagating shock occurs in a material under high strain rate loading conditions such as a high speed impact and can be fully characterized by the continuum governing equations. 
        However, it is well-known that a material's response to a given load is linked to effects at lower length scales. 
        This is especially true for damage, fracture, and plastic deformation induced by shock waves as such phenomena are the result of complex behavior at the microstructure. 
        Examples of such microstructural events include scattering, grain rotations, pore collapse, phase transformations, dislocations, void generation, and grain crushing \cite{RustyGray2012,Fensin2014,Bingert2014}.
        Therefore, in order to understand shock waves at the macroscale, one must also be able to describe and model them at the microscale.
        
        Non-equilibrium Molecular Dynamics (NEMD) simulations have been performed since the 1960s to study such \textit{atomistic} shock wave propagation \cite{holian1995atomistic}.
        In the past two decades, these NEMD simulations have been expanded to domains consisting of millions of atoms $(\sim O(10^5 - 10^9))$ and used to model increasingly complex events such as dislocation generation \cite{germann2004dislocation,Tramontina2017Simulation,righi2021towards,zhu2021collapse}, void nucleation \cite{bringa2010void,bisht2019investigation,tian2021anisotropic}, twinning \cite{Higginbotham2013Molecular,wu2021unveiling,zhu2021novel}, and shock-induced spallation \cite{Srinivasan2007,Fensin2014Effect,wang2021spall,chen2021molecular,dewapriya2021molecular}. 
        However, NEMD shock simulations can suffer from, among other things, wave reflections due to limited domain sizes which drastically reduce the total runtime. 
        To overcome these drawbacks, alternative atomistic shock wave frameworks have recently been developed, and some of the more popular of these formulations include the uniaxial Hugoniostat \cite{maillet2000uniaxial,maillet2002uniaxial}, the multiscale shock technique (MSST) \cite{Reed2003,reed2006analysis}, and the moving window method \cite{Zhakhovskii1997,davis2020one}.
        
        While contemporary atomistic methods have been very successful in modeling shock wave propagation and characterizing how defects influence shock behavior, even the largest benchmark simulations fall short of capturing continuum-level phenomena such as shock wave/boundary layer interactions and the resulting scattered elastic waves. 
        This is because limited computational resources restrict the total number of atoms that can realistically be incorporated into a microscale framework.
        Although computer processing power increases over time, physical limitations on computer architecture imply that simulations which model continuum events using purely atomistic techniques may never be possible.
        Therefore, a \textit{multiscale} scheme is needed which would be able to track a moving shock wave for a long time over a large domain as well as capture a shock's interaction with phonons and microstructural interfaces.
        Such a framework would locally retain atomistic information around a small region of interest (i.e. the shock wave front) and transition to continuum length scales throughout the rest of the domain. 
        
        Coupled atomistic-continuum (A-C) frameworks have been developed since the early 1990s to integrate the microscale and macroscale into a single computational model \cite{kohlhoff1991crack}.
        See \cite{mcdowell2020connecting,van2020roadmap,xiong2021multiscale,fish2021mesoscopic} for reviews on the current state of multiscale modeling in materials.
        Such methods can be roughly divided into two distinct categories: \textit{hierarchical} and \textit{concurrent} \cite{tadmor2011modeling}.
        Hierarchical methods presume a separation of spatial scales, and they formulate microscale information in terms of macroscopic behavior. 
        In other words, the lower-scale information is averaged and introduced into pure coarse-grained models in the form of constitutive equations.
        Some examples of hierarchical methods include multiple-level FEM \cite{miehe1999computational,lange2021efficient}, dislocation dynamics (DD) \cite{amodeo1990dislocation,zhang2021dislocation}, and the second gradient technique \cite{luscher2010second}.
        On the other hand, concurrent schemes connect the spatial scales such that the continuum region flanks or surrounds an inner atomistic region, and the two domains interact with each other directly.
        Concurrent methods are more applicable to shock wave simulations as they can describe large domains for a low computational cost while retaining important information at the wave front through atomistics.
        Hence, we use a concurrent scheme for the present formulation.
        
        One of the primary challenges with concurrent multiscale methods is ensuring compatibility at the interface of the fine-scaled and coarse-scaled regions so as to mitigate spurious wave reflections and ghost forces \cite{xu2018modeling}.
        Such non-physical phenomena arise for the following reasons: (i) a difference in governing equations exists between the atomistic and continuum regions, (ii) the spectrum of the coarse-scaled model has a much smaller cutoff frequency than that of the fine-scaled model causing the A-C interface to appear rigid to incoming high-frequency waves, and (iii) the interface region cannot support thermal vibrations of atoms. 
        Many concurrent schemes have been developed which address such issues in different ways \cite{tadmor2011modeling}.
        Some of these frameworks include the Coupling of Length Scales (CLS) method \cite{rudd1998coarse}, the Bridging Domain (BD) method \cite{xiao2004bridging}, the Coupled Atomistic Discrete Dislocation (CADD) method \cite{Shilkrot2002Coupled}, and the Quasicontinuum (QC) method \cite{tadmor1996quasicontinuum}.
        More recent concurrent frameworks use techniques such as the crystal plasticity finite element model (CPFEM) \cite{chakraborty2021concurrent} and microscale micromorphic Molecular Dynamics (MMMD) \cite{tong2020concurrent} to simulate crack propagation and dynamic fracture.
        Concurrent schemes such as these have had many achievements in modeling multiscale phenomena.
        Nevertheless, many of them involve a difference in material description across the interface, and most of them do not allow dislocation/phonon interactions or waves to travel into the continuum region \cite{xu2018modeling}.
        
        The Concurrent Atomistic-Continuum (CAC) method has been developed over the past decade to overcome some of the limitations in previous concurrent formulations \cite{xiong2011coarse,yang2013concurrent,xiong2014prediction,xiong2015concurrent,xu2016mesh,chen2017recent,chen2018passing,xu2018pycac,chen2019concurrent}.  
        Employing a unified multiscale framework built upon Atomistic Field Theory \cite{chen2005atomistic,chen2009reformulation}, CAC extends the Irving-Kirkwood method for connecting the atomistic and hydrodynamical equations \cite{irving1950statistical} to a two-level description of materials where the particle degrees of freedom are included within each primitive unit cell.
        In this way, CAC follows the solid state physics model of crystals whereby the structure is continuous at the lattice level but discrete at the atomic level, and a single set of governing equations is used throughout the entire domain \cite{chen2019concurrent}. 
        Hence, CAC not only supports dislocation/phonon interactions \cite{xiong2014sub,chen2017effects}, but it also allows dislocations and waves to pass from the atomistic region to the continuum region and vice versa \cite{xiong2011coarse,chen2018passing}. 
        While CAC has been very successful at modeling material defects and their motion and is actively being applied to study dislocation evolution and interactions \cite{xu2019comparison,xu2019sequential,li2019multiscale,selimov2021lattice}, it has not yet been extended to model shock wave propagation through a material.
        
        In the present work, we develop a long-time, moving window, A-C framework using CAC to model shock wave propagation through a one-dimensional monatomic chain.
        Specifically, we use the nonlinear Eulerian thermoelastic equations for shock compression of single crystals \cite{clayton2013nonlinear,clayton2014shock} to study the classic Riemann problem of a single propagating shock. 
        The shock is ``followed" for very long simulation times using a moving window technique as was accomplished previously in \cite{davis2020one} for a purely atomistic framework.
        In the present CAC formulation, the moving window is achieved in two distinct ways: (i) by tracking the shock in a conveyor fashion and (ii) by simultaneously refining the coarse-scaled region and coarsening the fine-scaled region at the speed at which the shock wave moves. 
        These methods allow us to model a propagating shock much longer than conventional NEMD simulations by focusing the shock front at the center of the fine-scaled region for the entire simulation.
        
        The paper is organized as follows.
        Section \ref{Sec: Problem Formulation} discusses the thermoelastic equations for shock wave propagation through anisotropic single crystals derived in \cite{clayton2013nonlinear} and outlines the problem statement.
        Section \ref{Sec: Computational Setup} describes the framework's geometry and boundary conditions and also presents the potentials and thermostats used in the simulations.
        Section \ref{Sec: CAC Method} reviews the governing equations and finite element implementation of CAC.
        Section \ref{Sec: Moving Window} discusses the two CAC moving window methods and how the shock is initialized in the domain. 
        Section \ref{Sec: Framework Verification} presents results from dispersion relation and phonon wave packet studies performed with the CAC framework.
        Section \ref{Sec: Shock Wave Results} presents results from the CAC moving window shock simulations and compares these to what is obtained from theory.
        Finally, Section \ref{Sec: Conclusion} concludes the paper and discusses future directions for this work.
        
    \section{Problem Formulation} \label{Sec: Problem Formulation}
        We consider an elastic monatomic chain with no defects under compression by an ideal one-dimensional (i.e. longitudinal) shock wave.
        Mathematically, we represent the shock as a propagating discontinuity across which there exists a jump in particle velocity, strain, and temperature.
        Material quantities ahead of the shock front have the superscript \textit{$-$}, and quantities behind the shock front have the superscript \textit{$+$}.
        In each simulation, particles ahead of the shock front are assumed to be at zero mean particle velocity, unstressed, unstrained, and at room temperature ($295$ K), and the shock propagates at a natural velocity $U_S$.
        The notation $\llbracket\cdot\rrbracket$ denotes the change in a given quantity $(\cdot)$ across the shock front.
        
        \subsection{Shock equations} \label{Sec: Shock equations}
            To characterize the shock wave at the continuum level, we use the nonlinear Eulerian thermoelastic shock equations derived in \cite{clayton2013nonlinear,clayton2014shock} for anisotropic crystals.
            Nonlinear elastic constitutive models are required when strains are relatively large.
            Since the experimental Hugoniot elastic limit typically occurs at small uniaxial compressive strains in ductile metals, predictions for material strength which omit plastic deformation are idealizations.
            However, applying analytical thermoelastic solutions to materials is beneficial for comparison with atomistic/multiscale simulations since these defect-free domains can be shocked to finite strains over short time scales and small volumes \cite{clayton2014shock,zimmerman2011elastic}.
            In this work, we conduct such simulations in a one-dimensional setting.
            
            For uniaxial loading, the `$11$' component of the deformation gradient behind the shock front is given as follows:
            \begin{equation}
                F_{11} = \frac{\partial x}{\partial X} = 1 + \frac{\partial u}{\partial X} = 1 + \epsilon = J
            \end{equation}
            where $u$ is the displacement, $\epsilon$ is the strain, and $J$ is the Jacobian determinant.
            We simulate compressive shocks, for which $0 < F \leq 1$ and $-1 < \epsilon \leq 0$, propagating with a positive velocity $U_S > 0$.
            Then, the only nonzero component of the Eulerian strain is
            \begin{equation}
                D = D_{11} = \frac{1}{2} \left(1 - F_{11}^{-2} \right) = \frac{1}{2} \left[1 - \frac{1}{(1 + \epsilon)^2} \right].
            \end{equation}
            Also, for uniaxial strain along the positive $x$-direction, we have the following:
            \begin{align} \label{Eq: ClaytonPressureAndJacobian}
                P &= -\sigma_{11} \nonumber \\
                J &= F_{11} = \left(1 - 2D \right)^{-1/2}
            \end{align}
            where $P$ is the axial shock stress.
            In this case, the term \textit{Eulerian} refers to a strain that is a function of the inverse deformation gradient and not necessarily one in spatial coordinates. 
            Therefore, the strain tensor $D$ is Eulerian but refers to material coordinates, so it can be used in simulations of anisotropic materials \cite{clayton2013nonlinear}. 
            As stated in \cite{birch1947finite}, the choice between Lagrangian and Eulerian is a matter of convenience, but the Eulerian formulation gives simpler expressions when modeling large compressions.
             
            It is well known that a planar shock wave propagating through an unstressed solid with velocity $U_S$ can be described by the Rankine-Hugoniot equations \cite{germain1973shock,thurston1974waves,davison2008fundamentals}
            \begin{align} \label{Eq: RankineHugoniotEquations}
                \llbracket P \rrbracket - \rho_0 U_s \llbracket v \rrbracket &= 0 \nonumber \\
                \llbracket v \rrbracket - U_s \llbracket 1-J \rrbracket &= 0 \nonumber \\
                \llbracket U \rrbracket - \frac{1}{2} \rho_0 \llbracket v^2 \rrbracket &= 0 
            \end{align} 
            where $\rho_0$, $v$, and $U$ denote density, particle velocity, and energy respectively. 
            As shown in \cite{clayton2013nonlinear}, analytical solutions to the planar shock problem can be derived when the material's internal energy is a linear function of entropy.
            Assuming uniaxial strain, the Eulerian fourth-order internal energy function and conjugate thermodynamic stress are given by \cite{clayton2014shock}
            \begin{equation} \label{Eq: ClaytonEulerianEnergy}
                \Hat{U} = \frac{1}{2}C_{11}D^2 + \frac{1}{6}\Hat{C}_{111}D^3 + \frac{1}{24} \Hat{C}_{1111}D^4 - \theta_0 \left(\Gamma_1 D + \frac{1}{2} \Hat{\Gamma}_{11}D^2 - 1 \right)\eta
            \end{equation}
            \begin{equation} \label{Eq: ClaytonEulerianStress}
                \Hat{S} = -J^3P = \frac{\partial \Hat{U}}{\partial D}.
            \end{equation}
            Here, $\theta_0 > 0$ and $\eta = 0$ are the respective temperature and entropy ahead of the shock front, $C_{11}$ is the second-order elastic constant, and $\Gamma_1$ is the first-order Gr\"{u}neisen parameter.
            The Eulerian third-order elastic constant, Eulerian fourth-order elastic constant, and Eulerian second-order Gr\"{u}neisen parameter are obtained by the relations \cite{weaver1976application,perrin1978application,clayton2013nonlinear}
            \begin{align}
                \Hat{C}_{111} &= C_{111} + 12C_{11} \label{Eq: ClaytonEulerianC111} \\
                \Hat{C}_{1111} &= C_{1111} - 18C_{111} - 318C_{11} \label{Eq: ClaytonEulerianC1111} \\
                \Hat{\Gamma}_{11} &= \Gamma_{11} + 4\Gamma_1 \label{Eq: ClaytonEulerianGruneisenConstant}
            \end{align}
            where $\Gamma_{11} = \Gamma_1$ assuming that $\rho \Gamma_{\beta} = \rho_0 \Gamma_{0 \beta} =$ constant \cite{wallace1980flow,clayton2013nonlinear}.
        
            Solving Eqs. 
            (\ref{Eq: ClaytonPressureAndJacobian}), 
            (\ref{Eq: RankineHugoniotEquations}), (\ref{Eq: ClaytonEulerianEnergy}), and (\ref{Eq: ClaytonEulerianStress}) simultaneously gives the following fifth-order polynomials for entropy $\eta$ generated across the shock front \cite{clayton2013nonlinear}:
            \begin{equation}
                \eta(D) = \sum_{k=0}^5 b_k D^k
            \end{equation}
            \begin{equation}
                b_0 = b_1 = b_2 = 0
            \end{equation}
            \begin{equation}
                b_3 = \frac{1}{12 \theta_0} \left(\hat{C}_{111} - 9C_{11} \right)
            \end{equation}
            \begin{equation}
                b_4 = \frac{1}{24 \theta_0} \left[\hat{C}_{1111} - 9\hat{C}_{111} - 6C_{11} + \Gamma_1 \left(\hat{C}_{111} - 9C_{11} \right) \right]
            \end{equation}
            \begin{equation}
                b_5 = \frac{1}{48 \theta_0} \left[-6\hat{C}_{1111} - 6\hat{C}_{111} - 9C_{11} + \Gamma_1 \left(\hat{C}_{1111} - 6 \hat{C}_{111} - 33C_{11} \right) + \Gamma_1^2 \left(\hat{C}_{111} - 9C_{11} \right) \right].
            \end{equation}
            Substituting these polynomials into Eq. (\ref{Eq: ClaytonEulerianEnergy}), we can obtain the following expression for the fifth-order conjugate stress (Eq. \ref{Eq: ClaytonEulerianStress}) \cite{clayton2013nonlinear}:
            \begin{equation}
                \hat{S} = C_{11}D + \frac{1}{2}\hat{C}_{111}D^2 + \left(\frac{1}{6}\hat{C}_{1111} - \theta_0 \Gamma_1 b_3 \right)D^3 - \theta_0 D^4 \left[ \left(\Gamma_1 b_4 + \hat{\Gamma}_{11} b_3 \right) + \left(\Gamma_1 b_5 + \hat{\Gamma} _{11} b_4 \right) D \right].
            \end{equation}
            From this conjugate stress, we can then use 
            Eqs. (\ref{Eq: ClaytonPressureAndJacobian}b) and
            (\ref{Eq: ClaytonEulerianStress}) to get an expression for the shock stress $P$:
            \begin{equation} \label{Eq: Eulerian Shock Pressure}
                P = -\left(1 - 2D \right)^{3/2} \hat{S}.
            \end{equation}
            Finally, the particle velocity $v$ and temperature $\theta$ behind the shock front as well as the shock velocity $U_S$ can be obtained from Eqs. (\ref{Eq: RankineHugoniotEquations}a), (\ref{Eq: RankineHugoniotEquations}b), (\ref{Eq: ClaytonEulerianEnergy}), and (\ref{Eq: Eulerian Shock Pressure}):
            \begin{equation} \label{Eq: ClaytonEulerianParticleVelocity}
                v = \biggl \{ \left(\frac{\hat{S}}{\rho_0} \right) \left[ \left(1-2D \right) - \left(1-2D \right)^{3/2} \right] \biggr \}^{1/2}
            \end{equation}
            \begin{equation} \label{Eq: ClaytonEulerianShockVelocity}
                U_S = v \left[1 - \left(1-2D \right)^{-1/2} \right]^{-1}
            \end{equation}
            \begin{equation} \label{Eq: ClaytonEulerianTemperature}
                \theta = \frac{\partial \hat{U}}{\partial \eta} = \theta_0 \left(1 - \Gamma_1 D - \frac{1}{2} \hat{\Gamma}_{11} D^2 \right).
            \end{equation}
            For every shock simulation, we use the third-order expression of Eqs. (\ref{Eq: ClaytonEulerianParticleVelocity}), (\ref{Eq: ClaytonEulerianShockVelocity}), and (\ref{Eq: ClaytonEulerianTemperature}).
            
        \subsection{Problem statement}
\begin{figure}[htpb]
                \centering
                \includegraphics[width=0.5\textwidth]{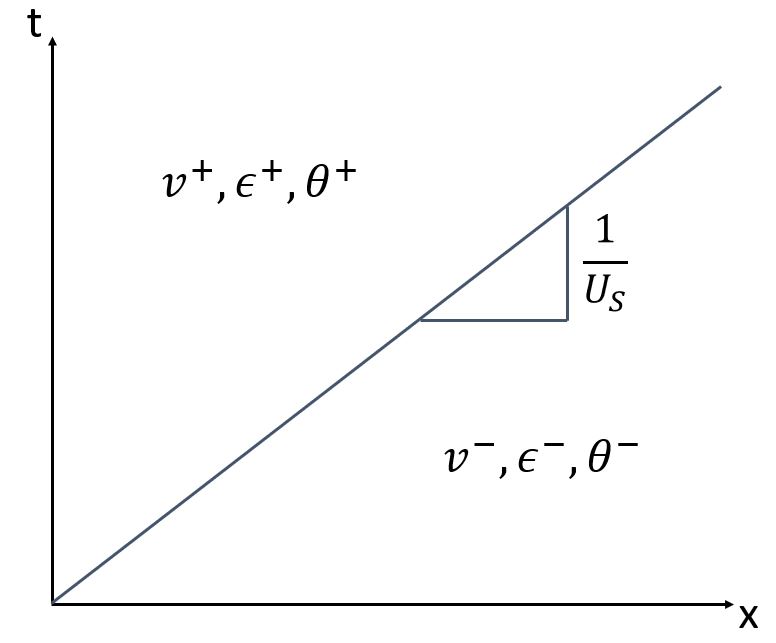}
                \caption{Riemann problem of a shock wave with constant states ahead of and behind the shock front.}
                \label{Fig:Riemann Shock}
            \end{figure}

            The material is described by the state variables $v$, $\epsilon$, and $\theta$ on either side of the shock front as well as the shock velocity $U_S$.
            Given an initial strain $\epsilon^-$, a final strain $\epsilon^+$, and state $(v^-,\theta^-)$ of the unshocked material, Eqs. (\ref{Eq: ClaytonEulerianParticleVelocity}), (\ref{Eq: ClaytonEulerianShockVelocity}), and (\ref{Eq: ClaytonEulerianTemperature}) can be used to compute $U_S$ and state $(v^+,\theta^+)$ of the shocked material.
            We incorporate these parameters into a moving window CAC framework to simulate long-time shock wave propagation given continuum shock states ahead of and behind the shock front. 
            Specifically, we study the classic Riemann problem of a single shock wave with constant states on either side as shown in Fig. \ref{Fig:Riemann Shock}.
        
    \section{Computational setup} \label{Sec: Computational Setup}
    
        \subsection{Geometry and boundary conditions} \label{Sec: Geometry and boundary conditions}
\begin{figure}[htpb]
                \centering
                \includegraphics[width=0.9\textwidth]{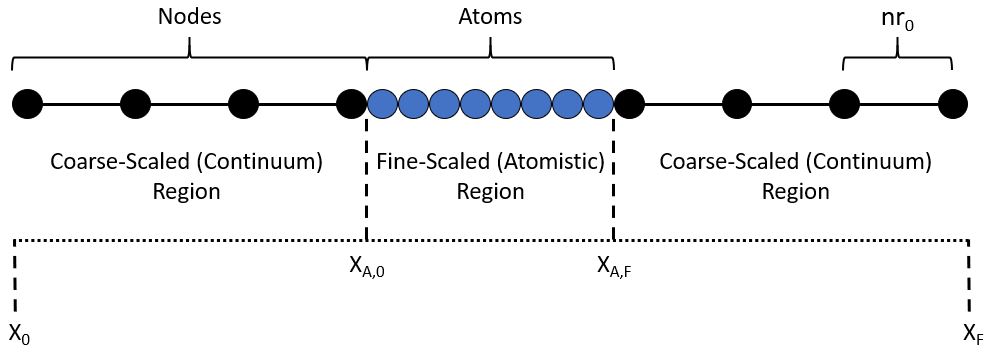}
                \caption{Schematic of the CAC domain.}
                \label{Fig:CACSystem}
            \end{figure}

            The one-dimensional CAC framework is implemented using an in-house C++ code. 
            The monatomic chain consists of $N$ particles which are split into three regions as seen in Fig. \ref{Fig:CACSystem}. 
            The particles in each coarse-scaled (continuum) region are separated by a distance of $nr_0$ and are referred to as \textit{nodes} in the present work. 
            Here, $n$ is some positive integer, and $r_0$ is the equilibrium spacing determined by the potential function. 
            These two coarse-scaled regions flank the inner fine-scaled (atomistic) region on either side.
            The particles in the fine-scaled region are separated by a distance of $r_0$ and are referred to as \textit{atoms} in the present work.
            Because CAC produces a unified atomistic-continuum framework using a single set of governing equations, the atoms and nodes have identical properties with the only difference being their inter-particle spacing. 
            As a result, the particles at the atomistic-continuum interfaces ($x_{A,0}$ and $x_{A,F}$) interact with each other directly while generating minimal ghost forces.
            
            For verification tests and non-shock simulations, we employ either standard fixed or periodic boundary conditions. 
            However, when modeling a propagating shock wave, the two coarse-scaled regions have distinct particle velocities, strains, and temperatures. 
            Therefore, to avoid introducing non-physical artifacts into the domain, a semi-periodic boundary condition method is employed during shock simulations.
            Specifically, the nodes at the ends of the chain ($x_0$ and $x_F$) are made neighbors with the nodes at the interfaces ($x_{A,0}$ and $x_{A,F}$ respectively). 
            The atoms and nodes at the A-C interfaces interact with each other directly as in non-shock simulations. 
        
        \subsection{Integration algorithm} \label{Sec: Integration algorithm}
            The CAC governing equation (Eq. \ref{Eq: FEM Governing Equation 2}) is a second order ordinary differential equation in time, and we solve it using the velocity Verlet algorithm \cite{swope1982computer} as seen below for 1D:
            \begin{align} \label{VelocityVerlet}
                x_{i} \left(t + \Delta t \right) &= x_i \left(t \right) + v_{i} \left(t \right) \Delta t + \frac{f_{i} \left(t \right)}{2m} \Delta t^2 \nonumber \\
                v_{i} \left(t + \frac{\Delta t}{2} \right) &= v_{i} \left(t \right) + \frac{\Delta t}{2} \frac{f_{i} \left (t \right)}{m} \nonumber \\
                f_{i} \left(t + \Delta t \right) &= f_{i} \left[x_{i} \left(t + \Delta t \right) \right] \nonumber \\
                v_{i} \left(t + \Delta t \right) &= v_{i} \left(t + \frac{\Delta t}{2}\right) + \frac{\Delta t}{2} \frac{f_{i} \left (t  + \Delta t \right)}{m}.
            \end{align}
            Here, $x_{i}$, $v_{i}$, and $f_{i}$ denote the position of the $i^{th}$ particle, its velocity, and the net force acting on it respectively.
            Additionally, $m$ is the atomic mass of the particle.
            The time step used in the integration algorithm is chosen to be $\Delta t = 0.001$ ps in order to minimize numerical error. 
            The velocity Verlet algorithm is adapted in the presence of the Langevin thermostat as explained in Sec. \ref{Sec: Thermostats}.
        
        \subsection{Interatomic potential and material parameters} \label{Sec: Interatomic potentials}
            We use the modified Morse interatomic potential function to calculate the integrand of the internal force density (Eq. \ref{Eq: Internal Force Density}).
            The standard Morse potential was modified by \cite{macdonald1981thermodynamic} to improve the agreement with experimental values for the thermal expansion of materials.
            The modified Morse potential only considers first nearest neighbor interactions and is given by the following expression \cite{macdonald1981thermodynamic}:
            \begin{equation} \label{Eq: Morse}
                \Pi(r_{ij}) = \frac{D_0}{2B-1}\left[e^{-2 \alpha \sqrt{B} (r_{ij} - r_0)} - 2Be^{-\alpha (r_{ij} - r_0) / \sqrt{B}}\right]
            \end{equation}
            where $r_{ij} = |x_i - x_j|$ is the magnitude of the displacement between particle $i$ and $j$, and $r_0$ is the distance at which the potential reaches the minimum.
            We perform shock simulations with the following FCC metals: Cu, Al, Ag, and Ni. 
            The parameters for these materials are given in Table \ref{Table: MorsePotentialParameters} where we note that each $r_0$ is equivalent to the equilibrium spacing along the [110] lattice direction of that particular element. 
            \begin{table}[h]
                \centering
                \caption{Material constants and Morse parameters of four different FCC metals \cite{macdonald1981thermodynamic}.}
                \label{Table: MorsePotentialParameters}
                \begin{tabular}{||c  c  c  c  c  c  c  c||}
                \hline
                \textit{Element} & \textit{mass (u)} & \textit{$\rho_0$ (g/$cm^3$)} & 
                $\Gamma_1$ & \textit{$r_0$ (\AA)} & \textit{$\alpha$ (\AA$^{-1}$)} & \textit{$D_0$ (eV)} & \textit{B} \\
                \hline
                Cu & 63.55 & 8.96 & 5.5486 & 2.5471 & 1.1857 & 0.5869 & 2.265 \\
                Al & 26.98 & 2.70 & 6.2753 & 2.8485 & 1.1611 & 0.3976 & 2.5 \\
                Ag & 107.87 & 10.49 & 5.9773 & 2.8765 & 1.1255 & 0.4915 & 2.3 \\
                Ni & 58.69 & 8.90 & 6.4699 & 2.4849 & 1.3909 & 0.6144 & 2.4 \\
                \hline 
                \end{tabular}
            \end{table}
            
            The expression for the Gr\"{u}neisen constant $\Gamma_1$ of a perfect crystal with pair interactions in \textit{d}-dimensional space is given as follows \cite{krivtsov2011derivation}:
            \begin{equation}
                \Gamma_1 = -\frac{1}{2d} \frac{\Pi'''(r_0)r_0^2 + (d-1)\left[\Pi''(r_0)r_0 - \Pi'(r_0) \right]}{\Pi''(r_0)r_0 + (d-1)\Pi'(r_0)}
            \end{equation}
            where $\Pi$ is the interatomic potential.
            For a one-dimensional chain, this equation reduces to
            \begin{equation} \label{Eq: Gruneisen constant 1D}
                \Gamma_1 = -\frac{1}{2} \frac{\Pi'''(r_0)r_0^2}{\Pi''(r_0)r_0}.
            \end{equation}
            We use Eq. (\ref{Eq: Gruneisen constant 1D}) to obtain the Gr\"{u}neisen constants given in Table \ref{Table: MorsePotentialParameters}.
            As seen, the $\Gamma_1$ values for a 1D chain are about 3x the experimental constants of a 3D crystal. 
            Without loss of generality, the CAC verification studies in Sec. \ref{Sec: Framework Verification} and the Appendix are performed using only the parameters for Cu.
            
        \subsection{Thermostat} \label{Sec: Thermostats}
            We impose and maintain temperature in the domain using the Langevin thermostat. 
            The Langevin thermostat is stochastic and thus adds a random force to the particle motion along with a damping term $\zeta$.  
            The one-dimensional equations of motion of this thermostat for a particle \textit{i} are as follows:
            \begin{align}\label{Langevin}
                f_{i}^{tot} \left(t \right) &= f_{i} \left(t \right) - \zeta m v_{i} \left(t \right) + \sqrt{\frac{2 k_{B} \theta \zeta m}{\Delta t}} \tilde{h_{i}} \left(t \right) \nonumber \\
                \left<\tilde{h}_{i} \left(t \right) \right> &= 0 \nonumber \\
                \left<\tilde{h}_{i, \alpha} \tilde{h}_{i, \beta} \left(t \right) \right> &= \delta_{\alpha \beta}
            \end{align}
            where $\alpha$ and $\beta$ denote Cartesian components, $k_{B}$ is Boltzmann's constant, and $\tilde{h}_{i}$ is a Gaussian random variable with a mean of zero and a variance of one. 
            Since Langevin is local in nature, the target temperatures $\theta^+$ and $\theta^-$ are specified for each atom. 
            We modify the velocity Verlet algorithm in the presence of the Langevin thermostat by performing the discretization used in LAMMPS \cite{schneider1978molecular}:
            \begin{align}
                v_i \left(t + \frac{\Delta t}{2} \right) &= v_i(t) - \frac{\Delta t}{2} \left[\frac{\nabla_i \Pi(t)}{m} + \zeta v_i(t) \right] + \sqrt{\frac{\Delta t k_B \theta \zeta}{m}}\tilde{h}_i 
                \nonumber \\
                x_i(t + \Delta t) &= x_i(t) + v_i \left(t + \frac{\Delta t}{2} \right) \Delta t
                \nonumber \\
                v_i \left(t + \Delta t \right) &= v_i \left(t + \frac{\Delta t}{2} \right) - \frac{\Delta t}{2} \left[\frac{\nabla_i \Pi(t + \Delta t)}{m} + \zeta v_i \left(t + \frac{\Delta t}{2} \right) \right] + \sqrt{\frac{\Delta t k_B \theta \zeta}{m}}\tilde{h}_i.
            \end{align}
            We note that because of the Verlet scheme, the time step in each velocity update is now $\frac{\Delta t}{2}$ rather than $\Delta t$.
            As per Langevin's requirements, we generate a different random variable for each particle during each velocity update. 
            
    \section{CAC Method} \label{Sec: CAC Method}
        The CAC method extends the Irving-Kirkwood non-equilibrium statistical mechanical formulation for hydrodynamics \cite{irving1950statistical} to a concurrent atomistic-continuum, two-level description of crystalline materials \cite{chen2005atomistic,chen2009reformulation,xiong2011coarse}.
        Specifically, CAC describes a crystalline system as a continuous collection of material points or \textit{unit cells}, and the domain is discretized with finite elements such that each element contains a collection of these primitive unit cells.
        Furthermore, each unit cell is embedded with a group of discrete particles, so the continuum lattice is directly linked to underlying atomistic behavior \cite{xiong2012concurrent,chen2018passing}.
        This two-level material description produces multiscale balance laws which characterize both the macroscopic lattice deformation as well as the microscopic rearrangement of particles \cite{chen2019concurrent}.
        As a result, a single set of governing equations regulates both the fine-scaled and coarse-scaled regions -- a unique feature of CAC \cite{chen2011assessment}.
            
        In this section, we present a very brief overview of the CAC formulation. 
        More details on the CAC framework and Atomistic Field Theory can be found in \cite{xiong2009multiscale,deng2010coarse,xiong2011coarse,chen2019concurrent}.
        The finite element form of the CAC governing equation is given as follows:
        \begin{equation} \label{Eq: FEM Governing Equation 2} 
            \rho^{\alpha} \Ddot{\textbf{u}}^{\alpha}(\textbf{x}) - \textbf{f}_{int}^{\alpha}(\textbf{x}) = \textbf{0}
        \end{equation}
        which is simplified to only be dependent on the internal force density (the effects of external forces are ignored).
        Here, \textbf{x} is the physical space coordinate, $\rho^{\alpha}$ is the local density of mass, $\textbf{u}^{\alpha}(\textbf{x})$ is the displacement of the $\alpha^{th}$ particle in the unit cell, and $\textbf{f}_{int}^{\alpha}(\textbf{x})$ is the internal force density generally expressed as
        \begin{equation} \label{Eq: Internal Force Density}
            \textbf{f}_{int}^{\alpha}(\textbf{x}) = \int_{\Omega(\textbf{x}')} \sum_{\beta = 1}^{N_a} \textbf{f} \left[\textbf{u}^{\alpha}(\textbf{x}) - \textbf{u}^{\beta}(\textbf{x}') \right] d\textbf{x}'
        \end{equation}
        where $\Omega(\textbf{x}')$ is the domain of $\textbf{x}'$, and $N_a$ is the total number of particles within each unit cell.
        Hence, the internal force density is a nonlinear, nonlocal function of relative displacements between neighboring particles within a given cutoff radius, and it can be obtained exclusively from the interatomic potential function \cite{yang2014concurrent}.
        At the lattice level, we use interpolation within an element to approximate the displacement field as follows \cite{xiong2011coarse}:
        \begin{equation} \label{Eq: Approximate Displacement Field}
            \Hat{\textbf{u}}^{\alpha}(\textbf{x}) = \boldsymbol{\Phi}_{\xi}(\textbf{x}) \textbf{U}_{\xi}^{\alpha}.
        \end{equation}
        Here, $\Hat{\textbf{u}}^{\alpha}(\textbf{x})$ is the displacement field for the $\alpha^{th}$ particle within a given element, $\boldsymbol{\Phi}_{\xi}(\textbf{x})$ is the shape function (linear in this work), and $\textbf{U}_{\xi}^{\alpha}$ is the displacement of the $\alpha^{th}$ particle within the $\xi^{th}$ unit cell. 
        We let $\xi = 1, 2, ..., n$ where $n$ is the total number of unit cells in the element.

        Substituting Eqs. (\ref{Eq: Internal Force Density}) and (\ref{Eq: Approximate Displacement Field}) into Eq. (\ref{Eq: FEM Governing Equation 2}), we can then use the method of weighted residuals to obtain the weak form of the governing equation by multiplying Eq. (\ref{Eq: FEM Governing Equation 2}) with a weight function $\boldsymbol{\Phi}_{\eta}(\textbf{x})$ and integrating over the entire domain:
        \begin{equation} \label{Eq: Weak Form of Governing Equation 2}
            \int_{\Omega(\textbf{x})} \left[\rho^{\alpha} \boldsymbol{\Phi}_{\eta}(\textbf{x}) \boldsymbol{\Phi}_{\xi}(\textbf{x}) \right] d\textbf{x} \Ddot{\textbf{U}}_{\xi}^{\alpha} -
            \int_{\Omega(\textbf{x})} \boldsymbol{\Phi}_{\eta}(\textbf{x}) \int_{\Omega(\textbf{x}')} \sum_{\beta = 1}^{N_a} \textbf{f} \left[\boldsymbol{\Phi}_{\xi}(\textbf{x}) \textbf{U}_{\xi}^{\alpha} - \boldsymbol{\Phi}_{\xi}(\textbf{x}') \textbf{U}_{\xi}^{\beta} \right] d\textbf{x}' d\textbf{x} = \textbf{0}.
        \end{equation}
        We solve this governing equation using numerical integration techniques and then utilize the velocity Verlet algorithm described in Sec. \ref{Sec: Integration algorithm} to obtain the position, velocity, and acceleration of each particle in the system.
        This finite element implementation of CAC eliminates a majority of the degrees of freedom in the continuum regions.
        For critical subdomains where atomistic behavior is important, the finest mesh is used such that the element length is equal to the atomic equilibrium spacing $r_0$.
        
\begin{figure}[htpb]
            \centering
            \includegraphics[width=0.9\textwidth]{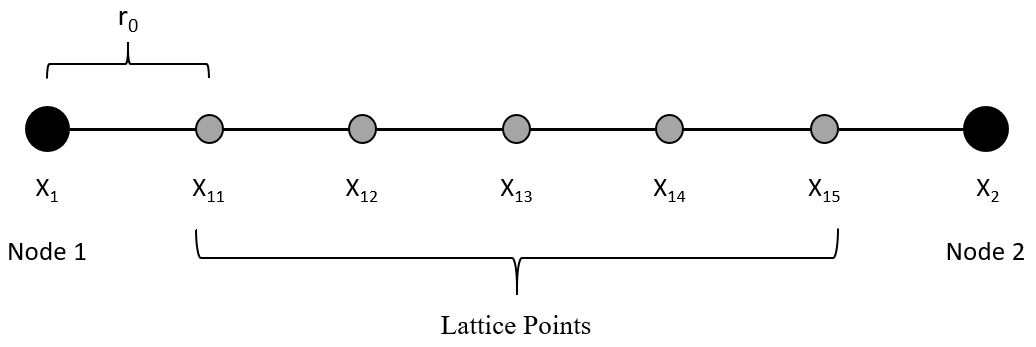}
            \caption{CAC coarse-scaled element for a one-dimensional monatomic chain.}
            \label{fig:LinearCACElement}
        \end{figure}

        We now specialize these equations for a one-dimensional monatomic chain using a 1D linear element in the coarse-scaled region as seen in Fig. \ref{fig:LinearCACElement}.
        For monatomic crystalline materials, the CAC primitive unit cell only contains one atom. 
        Hence, the ``unit cells" at positions $x_1$ and $x_2$ consist exclusively of nodes $1$ and $2$ respectively.
        Additionally, the element edge contains five lattice points whose positions are interpolated using Eq. (\ref{Eq: Approximate Displacement Field}) and which are excluded from the Verlet algorithm. 
        We do not apply any external forces and temperature is incorporated through the Langevin thermostat, so the governing equation reduces to the following in 1D:
        \begin{equation} \label{Eq: Matrix Form of Governing Equation in 1D}
            \textbf{M} \Ddot{\textbf{U}} - \textbf{F}^{int} = \textbf{0}
        \end{equation}
        where
        \begin{equation} \label{Eq: Mass Matrix in 1D}
            \textbf{M} = \int_{\Omega(x)} \left[\rho \boldsymbol{\Phi}(x) \boldsymbol{\Phi}(x) \right] dx
        \end{equation}
        \begin{equation} \label{Eq: Internal Force Density in 1D}
            \textbf{F}^{int} = \int_{\Omega(x)} \boldsymbol{\Phi}(x) \int_{\Omega(x')} \sum_{j = 1}^{n_{\alpha}} f \left[\boldsymbol{\Phi}(x) U_i - \boldsymbol{\Phi}(x') U_j \right] dx' dx.
        \end{equation}
        
        In Eq. (\ref{Eq: Matrix Form of Governing Equation in 1D}), $\textbf{M}$ is the complete mass matrix which is assembled from the lumped mass matrix (LMM) of each element in the domain.
        The standard LMM of a singular, two-node, one-dimensional element is defined as follows:
        \begin{equation} \label{Eq: Lumped Mass Matrix in 1D}
            \textbf{M}_{LMM}^e = \int_0^{L^e} \left[\rho^e \boldsymbol{\Phi}^e(x) \boldsymbol{\Phi}^e(x) \right] dx
            \approx \frac{m n^e}{2}
            \begin{bmatrix}
                1 & 0 \\
                0 & 1 \\
            \end{bmatrix}.
        \end{equation}
        Here, $m$ is the atomic mass, $n^e$ is the total number of particles and lattice points in the given element (where each particle gives $1/2$ contribution), $\rho^e$ is the atomic density of the element, $L^e$ is the element length, and $\boldsymbol{\Phi}^e(x)$ is the standard linear shape function in 1D.
        For the coarse-scaled element in Fig. \ref{fig:LinearCACElement}, the shape function at any position $x$ along the length of the element would be as follows:
        \begin{equation} \label{Eq: Linear Shape Function}
            \boldsymbol{\Phi}^e(x) = [\phi_1(x)\,\,\,\, \phi_2(x)] = \left[\frac{x_2 - x}{x_2 - x_1}\,\,\,\,\,\, \frac{x - x_1}{x_2 - x_1} \right] = \left[\frac{1 - \xi}{2}\,\,\,\,\,\, \frac{1 + \xi}{2} \right]
        \end{equation}
        where 
        \begin{equation}
            \xi = 2 \frac{x - x_C}{L^e},
        \end{equation}
        and $x_C$ is the central coordinate of the element.
        We note that in the natural coordinate system, $\xi = -1$ at node 1, $\xi = 0$ at the central point of the element, and $\xi = 1$ at node 2.  
        The coarse-scaled element in Fig. \ref{fig:LinearCACElement} has a length of $6r_0$ with two nodes and five lattice points, so its LMM is given as follows:
        \begin{equation}
            \textbf{M}_{LMM}^e \approx \frac{6m}{2}
            \begin{bmatrix}
                1 & 0 \\
                0 & 1 \\
            \end{bmatrix}
            = \begin{bmatrix}
                3m & 0 \\
                0 & 3m \\
            \end{bmatrix}.
        \end{equation}
        
        The terms $\Ddot{\textbf{U}}$ and $\textbf{F}^{int}$ are vectors of the respective accelerations and net internal forces for each particle in the chain. 
        In Eq. (\ref{Eq: Internal Force Density in 1D}), $n_{\alpha}$ is the number of neighbors of particle $i$ within a given cutoff radius, and $f$ is the force as a function of relative displacements acting on the particle. 
        If we let 
        \begin{equation} \label{Eq: 1D Force from Potential}
            f^{int}(x) = \int_{\Omega(x')} \sum_{j = 1}^{n} f \left[\boldsymbol{\Phi}(x) U_i - \boldsymbol{\Phi}(x') U_j \right] dx'
        \end{equation}
        then we can rewrite Eq. (\ref{Eq: Internal Force Density in 1D}) as follows:
        \begin{equation} \label{Eq: Simplified Internal Force Density in 1D}
            \textbf{F}^{int} = \int_{\Omega(x)} \boldsymbol{\Phi}(x) f^{int}(x) dx.
        \end{equation}
        A force $f^{int}(x)$ on particle $i$ at position $x$ is obtained exclusively from the interatomic potential function, and the corresponding net force is obtained through numerical integration.
        When finding the force $f^{int}(x)$ for a node in the coarse-scaled region, the surrounding lattice points are taken as neighbors and used to calculate relative displacements.
        
        We use nodal integration to calculate the internal force density in Eq. (\ref{Eq: Simplified Internal Force Density in 1D}).
        For a one-dimensional monatomic chain with many elements, the internal force vector would be assembled from the individual force vectors of each element.
        In a CAC framework, this internal force vector would contain net forces for both atoms and nodes.
        The only difference in the atomistic force calculations would be the element length $L^e$ (and thus $n^e$), as well as the fact that the neighbors of atoms are others atoms rather than lattice points. 
        As a result, lattice point positions would not have to be interpolated during the calculation of $f^{int}(x)$ for particles in the fine-scaled region.
            
    \section{Moving Window} \label{Sec: Moving Window}
        In traditional NEMD shock wave simulations, the shock cannot travel far before encountering a boundary, and this vastly limits the overall simulation time.
        In the present work, we circumvent this problem by utilizing a moving window.
        The moving window is similar in principle to moving boundary conditions used in \cite{holland1998ideal} and \cite{selinger2000dynamic} to model dynamic crack propagation.
        Incorporation of the moving window into shock simulations is inspired from \cite{Zhakhovskii1997} and \cite{zhakhovsky2011two}.
        In these works, a constant flux of material with a given density and velocity was fed into the simulation window by inserting a plane of atoms into the right boundary at regular time intervals. 
        With the CAC framework, we employ two distinct moving window methods to track the propagating shock and eliminate wave reflections: the \textit{conveyor} technique and the \textit{coarsen-refine} technique.

        \subsection{Geometry for a shock}\label{Sec: Geometry for a shock}
\begin{figure}[htpb]
                \centering
                \includegraphics[width=0.9\textwidth]{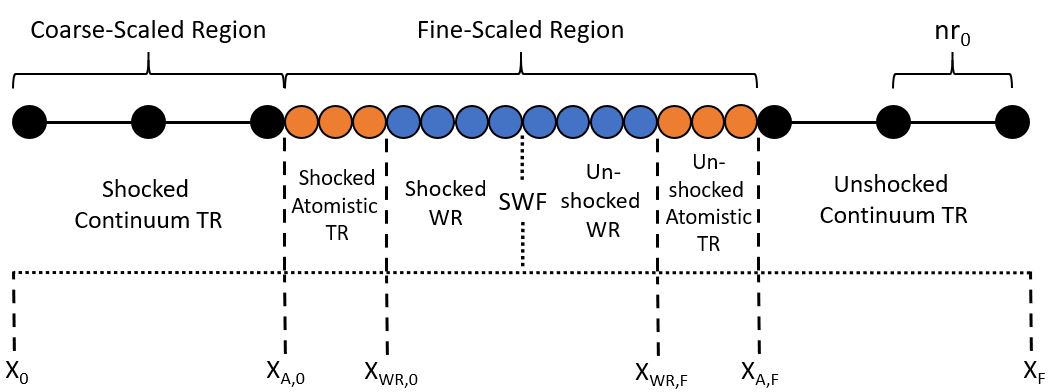}
                \caption{CAC geometry for moving window shock simulations.
                The black and orange circles represent the \textit{thermostat regions} (TRs) while the blue circles represent the \textit{window region} (WR).
                The shock wave front (SWF) originates at the center of the WR.}
                \label{Fig:CACMWGeometry}
            \end{figure}

            When performing shock wave simulations with the moving window, we utilize the entire CAC framework from Fig. \ref{Fig:CACSystem} and divide it into different regions as shown in Fig. \ref{Fig:CACMWGeometry}.
            The outer particles (black and orange circles) constitute the \textit{thermostat regions} (TRs), and these particles flank the interior atoms (blue circles) which constitute the \textit{window region} (WR). 
            We note that the two TRs encompass every continuum node as well as a small ``band" of fine-scaled atoms at each A-C interface in order for the WR to achieve the correct canonical ensemble \cite{qu2005finite}.
            All the particles to the left of the shock wave front (SWF) constitute the shocked material while all the particles to the right constitute the unshocked material.
            Semi-periodic boundary conditions are employed as detailed in Sec. \ref{Sec: Geometry and boundary conditions}, and the particles at the thermostat and A-C interfaces interact with each other directly.
            
            To minimize artificial kapitza resistance across the TR boundaries as well as efficiently absorb impinging transient waves, we specify the damping factor $\zeta$ in the Langevin thermostat to be a function of position relative to each TR/WR interface. 
            Specifically, we utilize the equation developed in \cite{qu2005finite} to linearly ramp damping in each atomistic TR as the distance from the TR/WR interface increases. 
            This equation is given as follows:
            \begin{equation}\label{LangevinDamping}
                \zeta = \zeta_{0} \left[1 - \frac{d \left(x_i \right)}{\textit{l}} \right]
            \end{equation}
            where $\zeta_{0}$ equals the maximum damping (one-half the Debye frequency), and $l$ is the length of the atomistic TR. 
            Here, $d(x_i)$ is the minimum absolute distance from atom $i$ at position $x$ to the respective A-C interface (either point $x_{A,0}$ or $x_{A,F}$). 
            Hence, for atoms in the fine-scaled TRs, the damping coefficient varies linearly from zero at the TR/WR interfaces to $\zeta_{0}$ at the A-C interfaces.
            The damping coefficient equals $\zeta_0$ throughout the coarse-scaled regions.
            This allows transient waves to enter the atomistic TRs and slowly be absorbed as they propagate to the continuum regions.
            Such a technique reduces spurious wave reflections and thus prevents artificial heating in the WR \cite{qu2005finite, miller2007hybrid}.
            
        \subsection{Conveyor technique} \label{Sec: Conveyor technique}
\begin{figure}[htpb]
                \centering
                \includegraphics[width=0.95\textwidth]{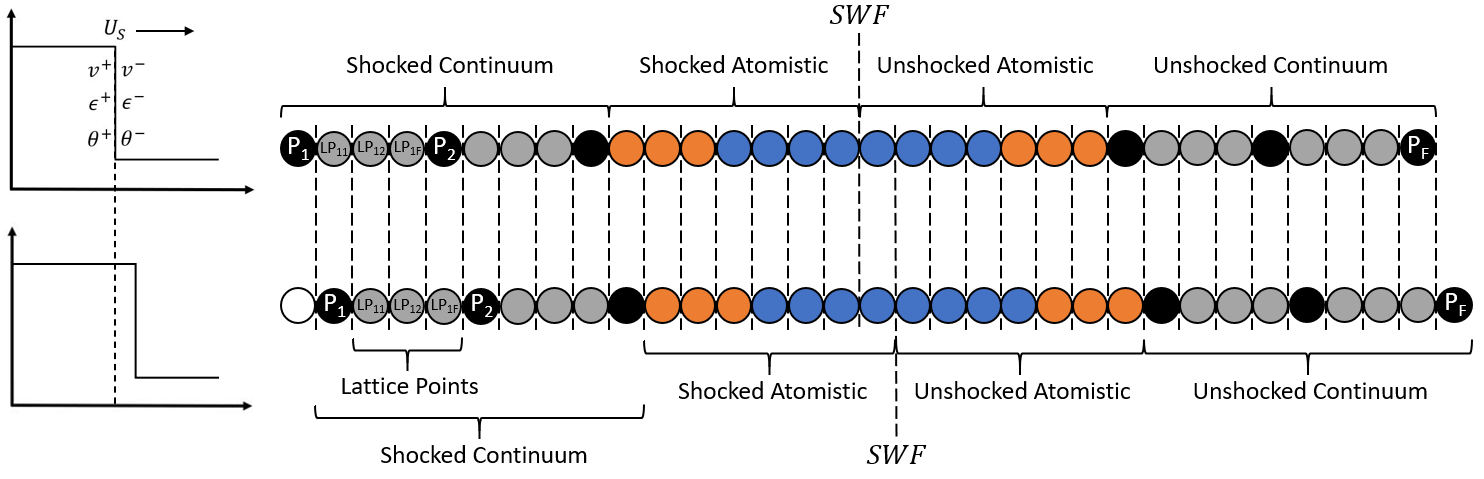}
                \caption{Moving window using the conveyor technique.}
                \label{Fig:CACMWConveyor}
            \end{figure}

            A schematic of the moving window using the conveyor technique is shown in Fig. \ref{Fig:CACMWConveyor}.
            The shock wave originates at the center of the WR as detailed in Sec. \ref{Sec: Initialization of the shock wave CAC} and immediately begins propagating forward into the unshocked material.
            After the shock has traveled one equilibrium spacing $r_0$, the atomic parameters (initial position, displacement, velocity, and acceleration) of the first coarse-scaled particle in the chain ($P_1$) are set equal to the parameters of the lattice point immediately to its right ($LP_{11}$).
            The initial position of $LP_{11}$ is stored, but its displacement, velocity, and acceleration must be interpolated using the linear shape function from Eq. (\ref{Eq: Linear Shape Function}) as shown in \cite{xu2016mesh}:
            \begin{align}
                U_{P_1} &\Rightarrow U_{LP_{11}} = \phi_{1}(x_{11})U_{P_1} + \phi_{2}(x_{11})U_{P_2} \nonumber \\
                \Dot{U}_{P_1} &\Rightarrow \Dot{U}_{LP_{11}} = \phi_{1}(x_{11})\Dot{U}_{P_1} + \phi_{2}(x_{11})\Dot{U}_{P_2} \nonumber \\
                \Ddot{U}_{P_1} &\Rightarrow \Ddot{U}_{LP_{11}} =  \phi_{1}(x_{11})\Ddot{U}_{P_1} + \phi_{2}(x_{11})\Ddot{U}_{P_2}
            \end{align}
            
            After this, the initial position of $LP_{11}$ is set equal to the initial position of $LP_{12}$ and so on throughout the first element until we reach the final lattice point in the element ($LP_{1F}$).
            Then, the initial position of $LP_{1F}$ is set equal to the initial position of $P_2$.
            For lattice points, only the initial positions are updated because their displacements are interpolated during the Verlet algorithm.
            These steps get repeated throughout each coarse-scaled element in the chain, and in the fine-scaled region, the parameters of a given atom are set equal to the parameters of its right neighbor without any interpolation.
            Lastly, we assign the following values to the final particle in the chain ($P_F$):
            \begin{align}
                X_{P_F} &= X_{P_{F}-1} + r_0 \nonumber \\
                U_{P_F} &= 0 \,\, \mathrm{km} \nonumber \\
                \Dot{U}_{P_F} &= 0 \,\, \mathrm{km/sec} \nonumber \\
                \Ddot{U}_{P_F} &= 0 \,\, \mathrm{km/sec^2}
            \end{align}
            where local atomic energy fluctuations induced near $x_F$ are damped by the Langevin thermostat as in \cite{zhakhovsky2011two}.
            
            This shifting process occurs iteratively with a frequency of $\tau^{-1} = U_{S} / r_0$ and effectively allows the monatomic chain to ``follow" the propagating shock wave.
            Therefore, if the simulated shock velocity matches the analytical shock velocity, the SWF should remain at the center of the WR throughout the entire simulation.
            We refer to this as a conveyor method because the chain moves in the positive x-direction with a constant velocity, and the coarse-scaled and fine-scaled regions always have the same respective lengths.
            Hence, no real refinement or coarsening of the domain takes place.
            
            Due to the specified shifting frequency, the highest time resolution in the conveyor technique is $r_0/U_S$. 
            This is solely because the region of interest in the domain is the SWF which travels with a known velocity of $U_S$.
            Phenomena with longer or shorter time scales would indeed be captured in the domain, but depending on their relative velocity, they may either outpace or fall behind the wave front and reflect off the A-C interfaces.
            The time resolution of the conveyor method could easily be increased by shifting the simulation cell more frequently.
            The only limiting factor would be the time step of the integration algorithm.
            Hence, the conveyor technique is highly adaptable to other dynamic behavior in materials such as crack growth or dislocation evolution.
            
        \subsection{Coarsen-refine technique} \label{Sec: Coarsen-refine technique}
\begin{figure}[htpb]
                \centering
                \includegraphics[width=0.95\textwidth]{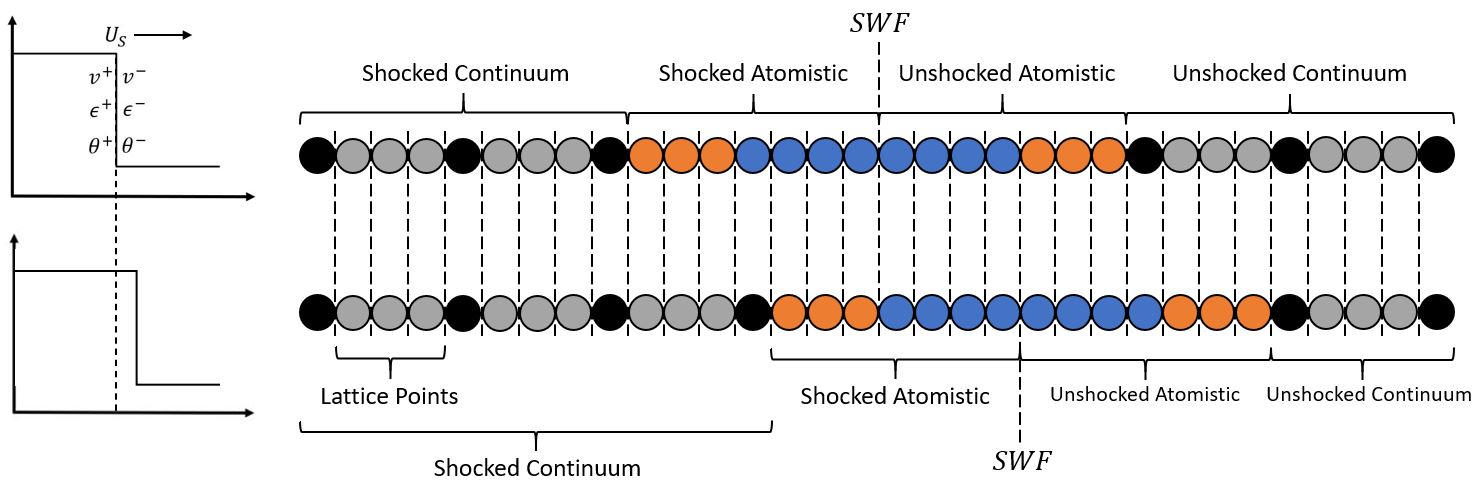}
                \caption{Moving window using the coarsen-refine technique.}
                \label{Fig:CACMWCoarsenRefine}
            \end{figure}

            A schematic of the moving window using the coarsen-refine technique is shown in Fig. \ref{Fig:CACMWCoarsenRefine}.
            As in the conveyor technique, the SWF originates at the center of the WR and propagates into the unshocked material.
            In the coarsen-refine method, however, atoms near the left A-C interface become lattice points while lattice points near the right A-C interface become atoms. 
            Therefore, the moving window update does not occur until the shock has traveled a distance of one element length $nr_0$.
            After this, the last node in the shocked continuum region (the node at the left A-C interface) becomes the second to last node in the shocked continuum region.
            Next, $n-1$ lattice points are assigned the initial positions of the corresponding $n-1$ atoms in the shocked atomistic region.
            Then, the first atom in the shocked atomistic region becomes the new A-C interface node and is assigned the initial position, displacement, velocity, and acceleration of the atom $n-1$ positions ahead of it.
            Effectively, a new continuum element has been created behind the SWF, and the shocked material has been coarsened. 
            
            This atomic parameter assignment occurs throughout the fine-scaled region as long as the current atom plus $n-1$ is less than or equal to the first node in the unshocked continuum region.
            After this point, each fine-scaled atom is assigned the parameters of the given lattice point $n-1$ positions ahead of it.
            As in the conveyor technique, the initial positions of the lattice points are stored, but their displacements, velocities, and accelerations must be interpolated.
            Finally, the second node in the unshocked continuum region maintains its parameters from the Verlet update, but it is now defined as the first node in the unshocked continuum region.
            Effectively, a continuum element ahead of the SWF has been transformed into atoms, and the unshocked material has been refined.
            After this entire process completes, the LMM is updated to reflect the modified mass distribution.
            
            The coarsen-refine technique occurs iteratively with a frequency of $\tau^{-1} = U_{S} / nr_0$, and it is distinct from the conveyor method because the initial and final positions of the framework do not change.
            Rather, the entire monatomic chain remains stationary and only the boundaries of the atomistic region are updated.
            We also note that the interface atoms of the atomistic TR bands are shifted accordingly.
            As a result, the shocked continuum region lengthens, the unshocked continuum region shortens, and the atomistic region tracks the SWF through the domain.
            For reasons similar to those mentioned in Sec. \ref{Sec: Conveyor technique}, the coarsen-refine technique introduces a time resolution of $nr_0/U_S$ into the domain due to the update frequency.
            Therefore, while phenomena such as elastic waves or dislocations could drift out of the fine-scaled region, they would merely reflect off the boundary or be absorbed by the TRs while the WR tracked the shock.
            
            The current coarsen-refine technique arises from a consideration of the balance between efficiency and accuracy.
            For example, merely refining the domain ahead of the wave front would result in total accuracy but no efficiency as the fine-scaled region would get larger and larger while the simulation evolved. 
            This would defeat the purpose of the CAC method and significantly increase the computational load over time.
            Likewise, merely coarsening the domain behind the wave front would result in total efficiency but no accuracy as the fine-scaled region would get smaller and smaller while the shock approached the A-C interface.
            Thus, we strike a balance between these two extremes by coarsening and refining equally-sized portions of the domain at the same rate.

        \subsection{Initialization of the shock wave} \label{Sec: Initialization of the shock wave CAC}
            We choose a final strain $\epsilon^+$ and use Eqs. (\ref{Eq: ClaytonEulerianParticleVelocity}) and (\ref{Eq: ClaytonEulerianShockVelocity}) to obtain the mean particle velocity $v^+$ and shock front velocity $U_S$. 
            This mean particle velocity represents a new equilibrium velocity for particles in the shocked region, and the integration algorithm is updated accordingly.
            The imposed strain causes the shocked region to experience uniaxial compression, and the particles obey the Cauchy-Born rule such that their new positions follow the overall strain of the material. 
            The non-zero particle velocity and compressive strain cause the shocked region to reach its final state and produce a forward-propagating shock wave beginning at the center of the WR. 
            The temperature rise $\theta^+$ is calculated from Eq. (\ref{Eq: ClaytonEulerianTemperature}) and imposed in the left TR, so the entire shocked region achieves this temperature \cite{qu2005finite}. 
            The fine-scaled TR bands are far enough away from the SWF (the non-equilibrium region) such that they are in regions of ``local" equilibrium.
            This ensures the validity of applying thermostats onto strained sections of the domain \cite{maillet2000uniaxial}.
    
    \section{Verification Studies} \label{Sec: Framework Verification}
        In this section, we present verification studies and phonon wave packet tests performed with the CAC framework.
        For the sake of brevity, results from the first two verification studies are presented and discussed in \ref{App: Force vs. displacement tests} and \ref{App: Temperature equilibration}.
        In \ref{App: Force vs. displacement tests}, we performed force vs. displacement tests to verify that Eq. (\ref{Eq: Internal Force Density in 1D}) was correctly implemented into the one-dimensional framework as well as ensure force matching at the A-C interfaces. 
        We found that the spring constants of the fine-scaled, coarse-scaled, and CAC domains were nearly identical.
        Next, in \ref{App: Temperature equilibration}, we performed temperature equilibration tests to ensure that the thermostat damping method discussed in Sec. \ref{Sec: Geometry for a shock} maintained the correct temperature in the undamped WR.
        We found that the Langevin thermostat equilibrated the system to the given input temperature and maintained this temperature for the entire runtime.
    
        \subsection{Predicting phonon dispersion relations} \label{Sec: Predicting phonon dispersion relations}
            We first verify that our CAC framework can produce the correct dispersion relation because such a relation characterizes the dynamics of a crystalline system and provides a direct test of a given theoretical model \cite{peckham1964phonon}.
            Traditionally, the dispersion relation of a crystalline system is obtained through Lattice Dynamics (LD). 
            Although LD provides a fundamental description of phonon properties in perfect crystals at low temperatures, it cannot account for anharmonic behavior like phonon-phonon scattering.
            Finite temperature effects can be incorporated into LD through quasi-harmonic methods, but such methods are typically unwieldy and impractical to implement \cite{rutledge1998comparison,turney2009predicting}.
            CAC naturally incorporates anharmonic phenomena, and fortunately, alternative methods to LD have been developed to evaluate phonon properties of crystalline systems in MD settings.
            The most common technique is the calculation of the phonon spectral energy density $\varphi(k,\omega)$ defined as the average kinetic energy per unit cell as a function of wavevector $k$ and angular frequency $\omega$. 
            In 1D, $\varphi(k,\omega)$ is given as \cite{thomas2010predicting}
            \begin{equation} \label{Eq: Spectral Energy Density 1D}
                \varphi(k,\omega) = \frac{m}{4 \pi \tau_0 N} \left| \int_0^{\tau_0} \left\{ \sum_{n=1}^{N} \Dot{u}_n(t) \times \mathrm{exp} \left[i k \cdot x_{n}(t_0) - i \omega t \right] \right\} dt \right|^2,
            \end{equation} 
            where $\tau_0$ is the total simulation time, $N$ is the total number of particles, $\Dot{u}_n(t)$ is the velocity of particle $n$ at time $t$, and $x_{n}(t_0)$ is the initial position of particle $n$.
            
            We use Eq. (\ref{Eq: Spectral Energy Density 1D}) to calculate the phonon  dispersion curve of the CAC framework by postprocessing atomic velocities in the fine-scaled region and nodal velocities in the coarse-scaled regions (Fig. \ref{Fig:CACSystem}).
            For this calculation, the monatomic chain contains 260 atoms in the fine-scaled region and 20 nodes in each coarse-scaled region for a total of 300 particles.
            The element length in the coarse-scaled regions is defined as $6r_0$, and standard periodic boundary conditions are applied at $x_0$ and $x_F$.
            
\begin{figure}[htpb]
                \centering
                \includegraphics[width=0.55\textwidth]{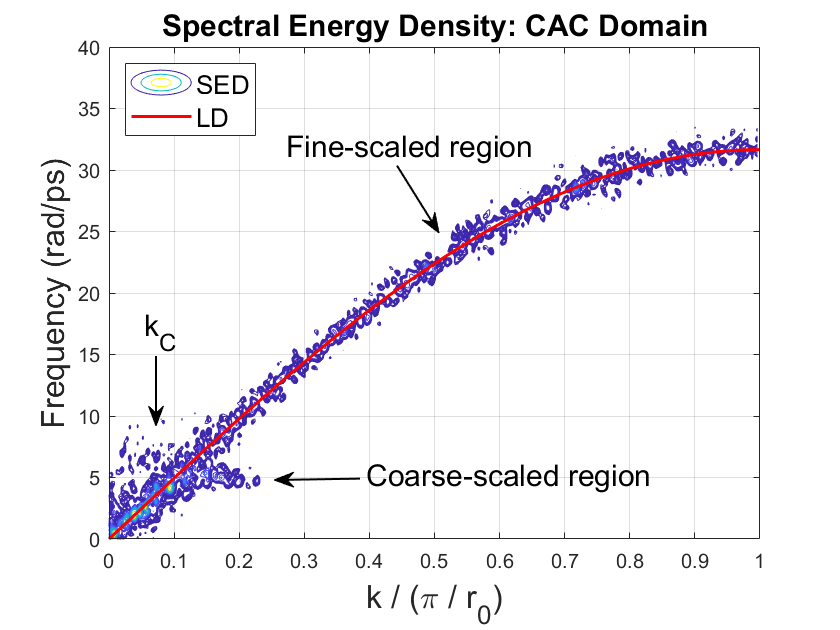}
                \caption{Phonon spectral energy density contour plot of a CAC monatomic chain calculated using the \textit{Langevin} thermostat.
                The red line represents the analytical dispersion relation obtained from LD, and the simulation was performed at $10$ K.}
                \label{Fig:SEDMorse}
            \end{figure}

            The spectral energy density plot as well as the dispersion relation obtained analytically through LD using the modified Morse potential can be seen in Fig. \ref{Fig:SEDMorse}.
            The dispersion relation for a one-dimensional monatomic crystal is given by the following standard equation:
            \begin{equation}
                \omega = \sqrt{\frac{4 C}{m}} \left|\mathrm{sin} \left(\frac{k r_0}{2} \right) \right|
            \end{equation}
            where $C$ is the elastic constant defined as the second derivative of the interatomic potential function at $r_0$ in this case.
            This simulation was performed at $10$ K.
            The frequency resolution on the vertical axis is $0.004$ rad/ps, the wavevector resolution on the horizontal axis is $0.004$, and the contours indicate the magnitude of the spectral energy density for each ($k$, $\omega$) combination.
            
            In Fig. \ref{Fig:SEDMorse}, we observe that the phonon dispersion relation obtained in the fine-scaled region of the CAC framework is identical to that obtained from LD.
            However, the dispersion relation for the coarse-scaled regions is only accurate for phonons whose wavevector is smaller than a critical value.
            The critical wavevector $k_C$ for an allowed error $\epsilon$ is calculated by the element length ($L = nr_0$) in the coarse-scaled regions according to the following equation \cite{chen2017ballistic}:
            \begin{equation}
                k_C = \underset{k}{\operatorname{max}} \left\{ \left| \mathrm{sin} \left(\frac{kr_0}{2} \right) - \mathrm{sin} \left(\frac{kL}{2} \right) \right| \le \epsilon \right\}.
            \end{equation}
            For this equation to be valid, the length values must be in nanometers, so $r_0 = 0.25471$ nm and $L = 6r_0 = 1.5283$ nm in our case for Cu.
            We choose an allowable error of $\epsilon = 5$\% which corresponds to a critical wavevector of $k_C = 0.064 \, \pi/r_0$, and a critical wavelength $\lambda_C = 2\pi/k_C = 7.96$ nm.
            These results are consistent with spectral energy density plots obtained in previous works which use the CAC method for phonon heat transport and the prediction of phonon properties \cite{xiong2014prediction,chen2017ballistic,chen2018passing}.
        
        \subsection{Phonon wave packet reflections at the A-C interface} \label{Sec: Wave packet reflections at the A-C interface}
            To visualize the transmission and reflection of waves at the A-C interface, we perform phonon wave packet studies using the CAC framework.
            We create the wave packet from a single branch of the dispersion relation obtained in Sec. \ref{Sec: Predicting phonon dispersion relations} with a narrow frequency range and well-defined polarization.
            Specifically, for each phonon mode, we know the corresponding wavevector $k$ and angular frequency $\omega$ from the LD dispersion relation or spectral energy density plot, and we can generate a Gaussian wave packet by assigning an initial displacement $U_n$ to the $n^{th}$ particle as follows \cite{schelling2002phonon,schelling2004kapitza,wei2019phonon}:
            \begin{equation} \label{Eq: Wave Packet Displacement at t = 0}
                U_n = A \epsilon(k) \mathrm{exp} \left[i k (x_n - x_0) \right] \mathrm{exp} \left[-(x_n - x_0)^2 / \xi^2 \right].
            \end{equation}
            In Eq. (\ref{Eq: Wave Packet Displacement at t = 0}), $A$ is the displacement amplitude, $\epsilon(k)$ is the polarization vector ($\epsilon(k) = 1$ in 1D), $x_n$ is the position of the $n^{th}$ particle, $x_0$ is the position of the particle at the center of the wave packet, and $\xi$ is the spatial extent of the wave packet.
            The time-dependent displacement and velocity are calculated as follows:
            \begin{equation}
                U_n^{*} = U_n \times e^{-i \omega t}
            \end{equation}
            \begin{equation}
                V_n^{*} = \frac{dU_n^{*}}{dt} = -i \omega U_n \times e^{-i \omega t},
            \end{equation}
            and at $t = 0$, the initial velocity of the wave packet is given as \cite{wei2019phonon}
            \begin{equation} \label{Eq: Wave Packet Velocity at t = 0}
                V_n = \omega \times Imag(U_n).
            \end{equation}
            After the wave packet is initialized in the atomistic region, it is allowed to propagate freely into the undisturbed medium, so only the displacement and velocity at $t = 0$ (Eqs. \ref{Eq: Wave Packet Displacement at t = 0} and \ref{Eq: Wave Packet Velocity at t = 0}) are used. 
        
\begin{figure}[htpb]
                \centering
                \begin{subfigure}{0.48\textwidth}
                    \includegraphics[width=\textwidth]{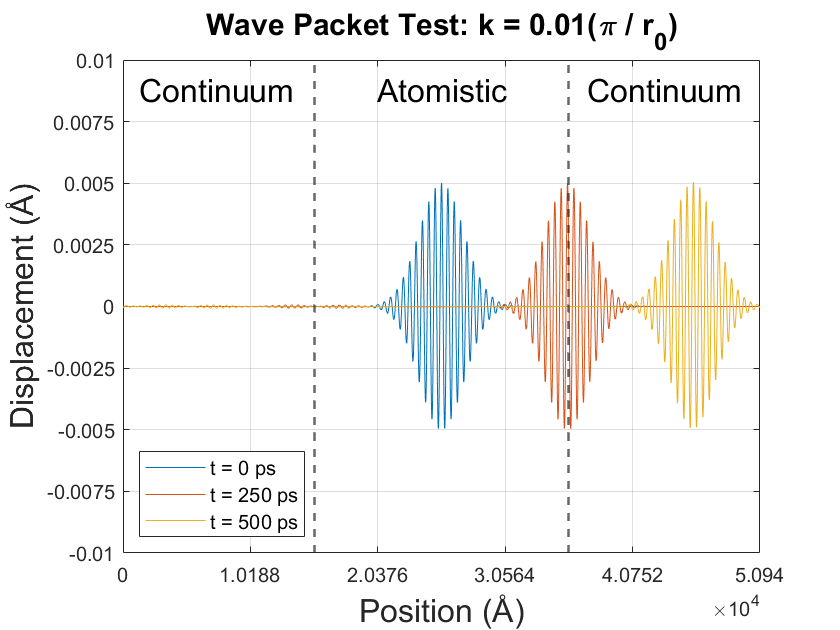}
                    \caption{}
                \end{subfigure}
                \begin{subfigure}{0.48\textwidth}
                    \includegraphics[width=\textwidth]{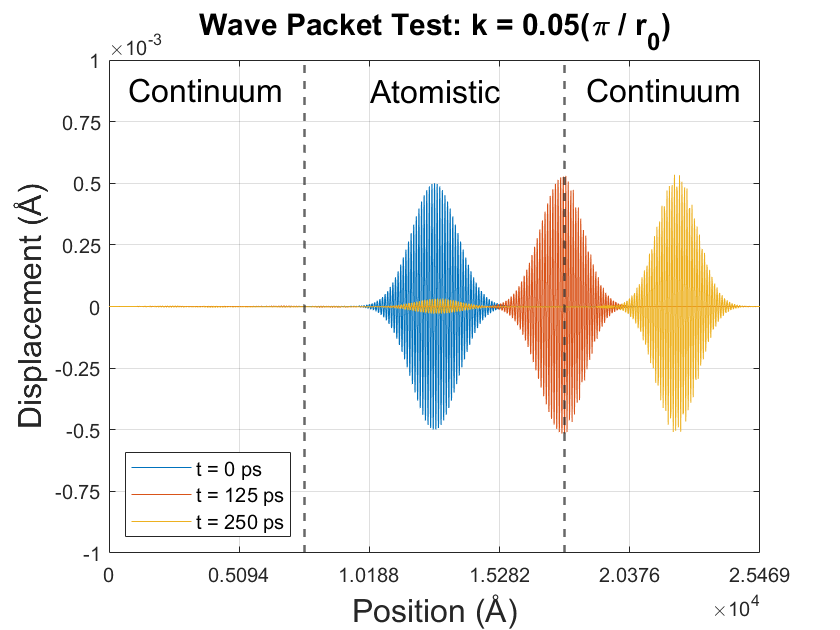}
                    \caption{}
                \end{subfigure}
                \\
                \begin{subfigure}{0.48\textwidth}
                    \includegraphics[width=\textwidth]{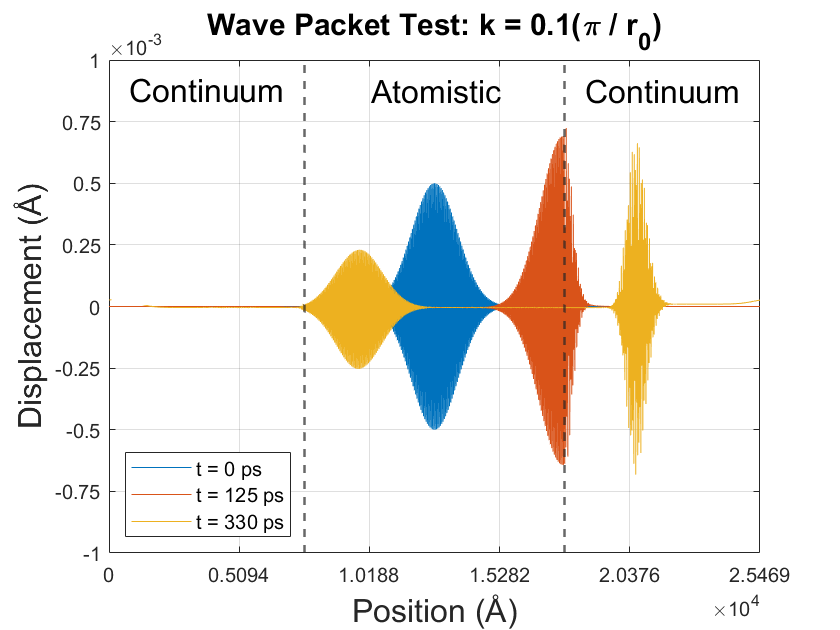}
                    \caption{}
                \end{subfigure}
                \begin{subfigure}{0.48\textwidth}
                    \includegraphics[width=\textwidth]{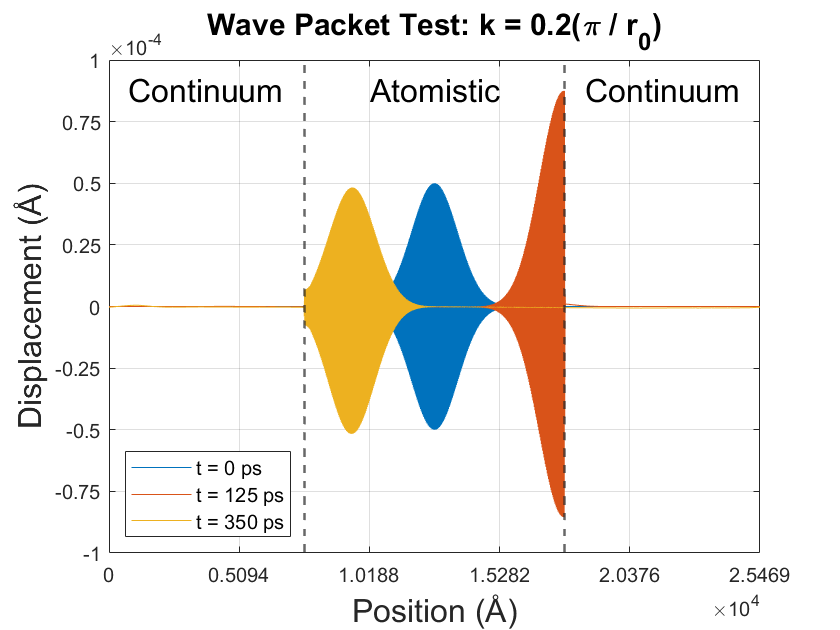}
                    \caption{}
                \end{subfigure}
                \caption{Phonon wave packet simulations performed with the following wavevectors: (a) $0.01 \, \pi/r_0$, (b) $0.05 \, \pi/r_0$, (c) $0.1 \, \pi/r_0$, and (d) $0.2 \, \pi/r_0$.}
                \label{Fig:WPTMorse}
            \end{figure}

            In Fig. \ref{Fig:WPTMorse}, we present four sets of CAC simulation results with wave packets of the following four central wavevectors: $k =$ $0.01$, $0.05$, $0.1$, and $0.2 \, \pi/r_0$.
            These wave packets are chosen to represent a range of values on the dispersion curve such that $k = 0.01 \, \pi/r_0$ is in the atomistic-continuum overlap region, $k = 0.05 \, \pi/r_0$ is slightly below the critical wavevector, $k = 0.1 \, \pi/r_0$ is slightly above the critical wavevector, and $k = 0.2 \, \pi/r_0$ is a short-wavelength phonon which cannot be modeled by the coarse-scaled region.
            We observe that the long-wavelength wave packet ($k = 0.01 \, \pi/r_0$) achieves nearly complete transmission across the A-C interface, and the $k = 0.05 \, \pi/r_0$ wave packet has only a small reflection.
            However, the shorter-wavelength wave packet ($k = 0.1 \, \pi/r_0$) has less transmission at the A-C interface, while the $k = 0.2 \, \pi/r_0$ wave packet is completely reflected.
            The energy transmission of wave packets at the A-C interface for various wavevectors can be seen in Table \ref{Table: WavePacketEnergyTransmission}.
            These phonon wave packet results confirm that the reflections at the A-C interface are a direct result of the numerical discrepancy between the atomistic and continuum regions as shown in previous studies \cite{xiong2014prediction,chen2018passing}.
            \begin{table}[htpb]
                \centering
                \caption{Effect of wavevector on the energy transmission of a phonon wave packet traveling from the fine-scaled region to the coarse-scaled region.}
                \begin{tabular}{||c  c||}
                \hline
                \textit{k / ($\pi / r_0$)} & \textit{$\%$ energy transmission} \\
                \hline
                0.01 & 99.97 \\
                0.05 & 99.61 \\
                0.06 & 99.11 \\
                0.07 & 98.10 \\
                0.08 & 95.83 \\
                0.09 & 90.63 \\
                0.10 & 73.35 \\
                0.20 & 0.00 \\
                \hline
                \end{tabular}
                \label{Table: WavePacketEnergyTransmission}
            \end{table}
        
        \subsection{Moving window with phonon wave packets} \label{Sec: Moving window with phonon wave packets}
            To verify that the two CAC moving window methods described in Secs. \ref{Sec: Conveyor technique} and \ref{Sec: Coarsen-refine technique} maintain a wave at the center of the fine-scaled region and do not produce any spurious phenomena at the A-C interfaces, we perform moving window simulations with a phonon wave packet. 
            We choose a medium-wavelength wave packet ($k = 0.2 \, \pi/r_0$) such that the wave would ordinarily be reflected at the A-C interface if allowed to propagate freely. 
            The only difference with these simulations is that the Langevin thermostat is not applied to the domain -- the particles are initially at rest.
            
\begin{figure}[htpb]
                \centering
                \begin{subfigure}{0.48\textwidth}
                    \includegraphics[width=\textwidth]{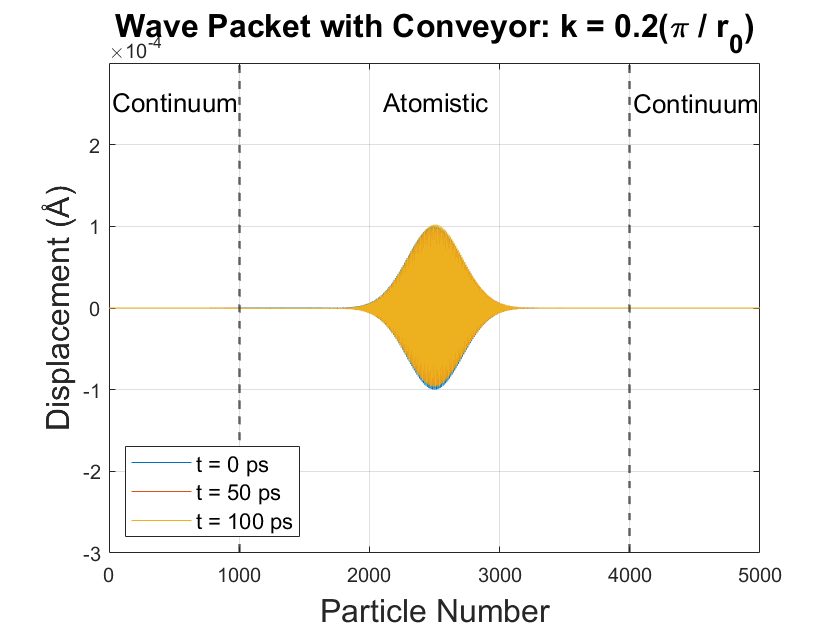}
                    \caption{}
                \end{subfigure}
                \begin{subfigure}{0.48\textwidth}
                    \includegraphics[width=\textwidth]{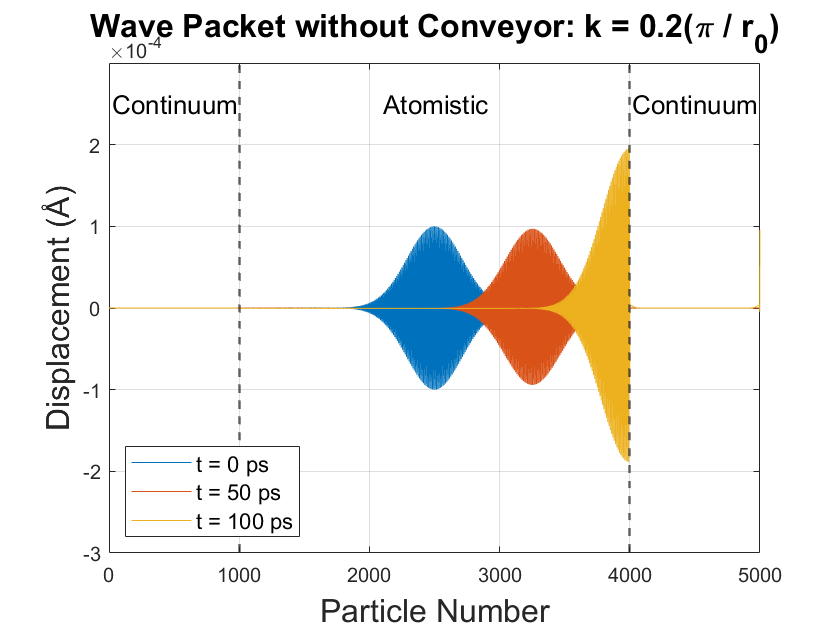}
                    \caption{}
                \end{subfigure}
                \caption{Wave packet simulations performed both (a) with and (b) without the conveyor method.}
                \label{Fig:WPT_MovingWindow}
            \end{figure}

            Simulation results using the medium-wavelength wave packet both with and without the conveyor mechanism are presented in Fig. \ref{Fig:WPT_MovingWindow}.
            We observe that this method effectively prevents the phonon wave packet from propagating forward, and the wave remains at the center of the atomistic region.
            Additionally, no spurious waves appeared at either A-C interface throughout the duration of the simulation.
            These results confirm that the conveyor moving window method works correctly and can track waves much longer than would be allowed in traditional MD simulations.
            
\begin{figure}[htpb]
                \centering
                \includegraphics[width=0.55\textwidth]{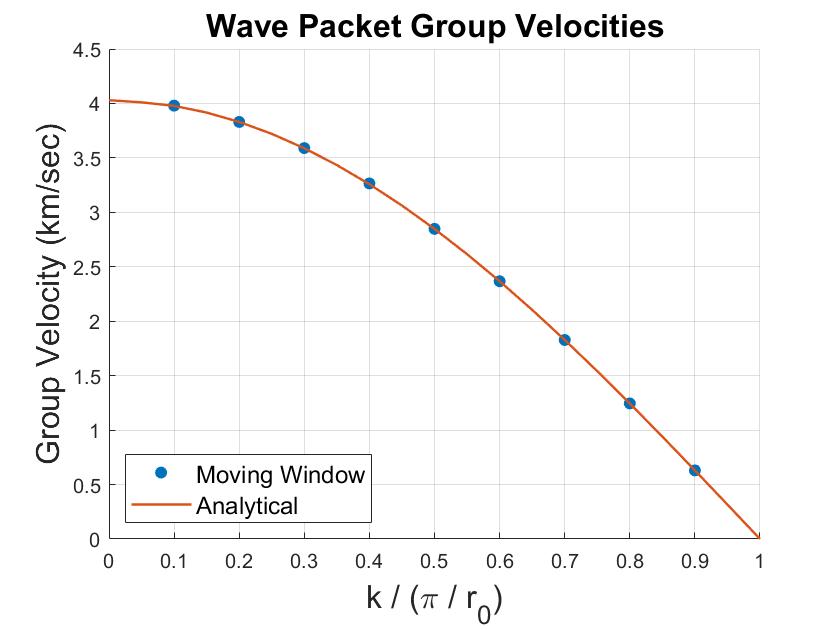}
                \caption{Analytical and simulated group velocities of phonon wave packets.}
                \label{Fig:WPT_GroupVelocity}
            \end{figure}

            When modeling a phonon wave packet with the conveyor method, the moving window update frequency is determined by the group velocity of the wave packet.
            In each simulation, we use the analytical group velocity given by $v_g = \frac{\partial \omega}{\partial k}$ to initialize the update frequency defined as $\tau^{-1} = v_g / r_0$.
            Then, we observe the drift in the propagating wave packet to calculate its actual group velocity.
            These analytical and simulated group velocities are plotted against the reduced wavevector in Fig. \ref{Fig:WPT_GroupVelocity}.
            Here, we see that the actual group velocity of the wave packet obtained from the conveyor method follows the analytical group velocity exactly over the full range of wavevectors.
            This confirms that the conveyor mechanism works correctly for both large and small wavelength waves and does not alter the wave speed in any nonphysical way.
                
\begin{figure}[htpb]
                \centering
                \includegraphics[width=0.55\textwidth]{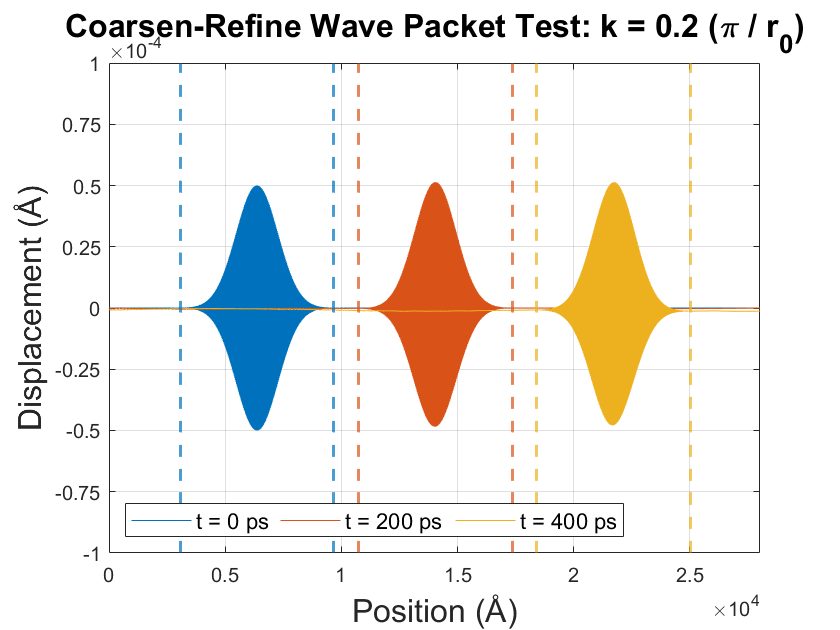}
                \caption{Time evolution of a phonon wave packet using the coarsen-refine technique.}
                \label{Fig:WPT_CoarseRefine}
            \end{figure}

            In Fig. \ref{Fig:WPT_CoarseRefine}, we present results of a wave packet simulation performed with the coarsen-refine technique.
            Here, the wave originates at the left of the domain and travels to the right over a period of 400 ps.
            The displacement profile of the wave packet at $0$ ps, $200$ ps, and $400$ ps is shown, and the dotted lines represent the position of the left and right A-C interface at each time. 
            In this case, the coarsen-refine update frequency is $\tau^{-1} = v_g / nr_0 = v_g / L$.
            We observe that the initial and final positions ($x_0$ and $x_F$) remain stationary while the atomistic region moves through the entire domain by the simultaneous coarsening of the left continuum region and refinement of the right continuum region.
            In effect, the left coarse-scaled region lengthens, and the right coarse-scaled region shortens as the simulation progresses in time.
            As a result, the fine-scaled region tracks the wave packet thus preventing it from ever encountering an interface. 
    
    \section{Shock Wave Results} \label{Sec: Shock Wave Results}
        In this section, we present results from shock wave simulations performed with both the conveyor and coarsen-refine moving window technique using the CAC domain from Fig. \ref{Fig:CACMWGeometry}.
        Before conducting these studies, we had to determine an appropriate length for the fine-scaled region to ensure that the structured shock front was fully-contained in this region and did not ``spill over" into the coarse-scaled regions.
        The length of the fine-scaled region is determined by the spatial shock thickness ($T_S$) which is calculated from the shock velocity ($U_S$) and rise time ($R_S$) by the following equation: $T_S = U_S \times R_S$.
        Studies have shown a wide range in the shock rise times for small-domain FCC metals with lower limits of $\sim 10^{-3}$ ns \cite{gahagan2000measurement} and upper limits of $\sim 10^0$ ns \cite{chhabildas1979rise} for the same material.
        To be safe, we choose a rise time of $10^0 = 1.0$ ns.
        The materials studied are Cu, Al, Ag, and Ni with a maximum compressive strain of $-0.1$.
        Using a rise time of 1.0 ns with the analytical shock velocity obtained from $\epsilon^+ = -0.1$, we get the following upper limits of the shock thickness for Cu, Al, Ag, and Ni respectively: 55,170 \AA \, ($\sim$21,660 atoms), 79,700 \AA \, ($\sim$27,980 atoms), 42,680 \AA \, ($\sim$14,838 atoms), and 70,700 \AA \, ($\sim$28,452 atoms).
        Therefore, without loss of generality and to be extra precautious, we perform all of our shock wave simulations with 40,000 atoms in the fine-scaled region.
        
        \subsection{Conveyor method}
\begin{figure}[htpb]
                \centering
                \begin{subfigure}{0.48\textwidth}
                    \includegraphics[width=\textwidth]{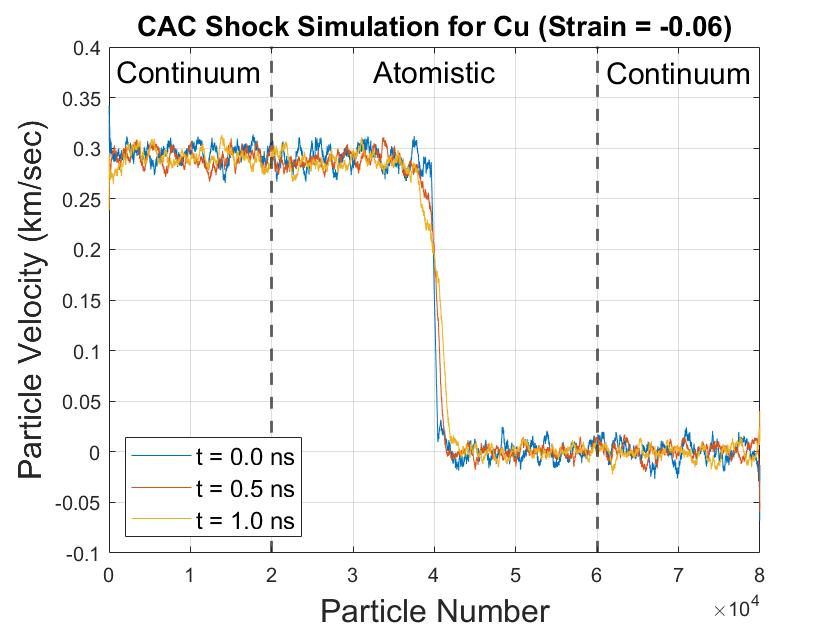}
                    \caption{}
                \end{subfigure}
                \begin{subfigure}{0.48\textwidth}
                    \includegraphics[width=\textwidth]{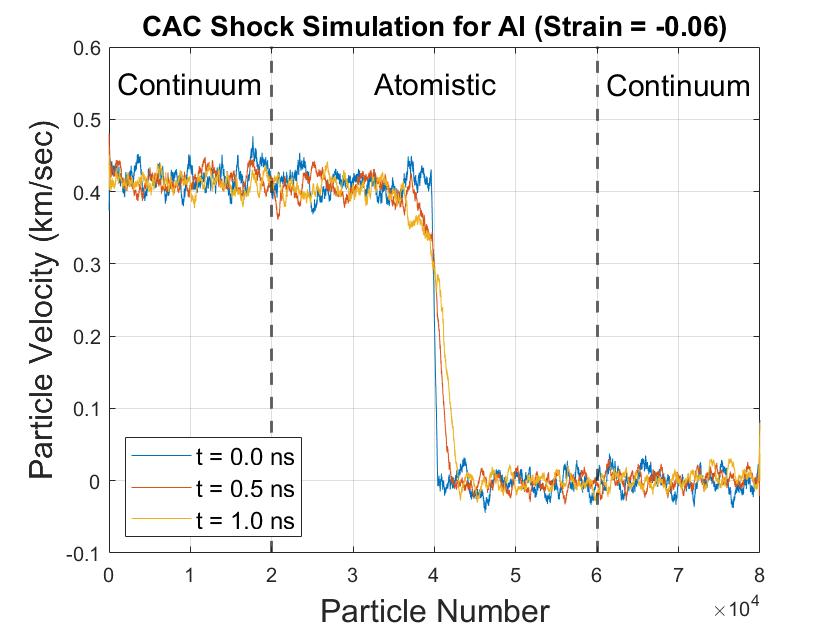}
                    \caption{}
                \end{subfigure}
                \\
                \begin{subfigure}{0.48\textwidth}
                    \includegraphics[width=\textwidth]{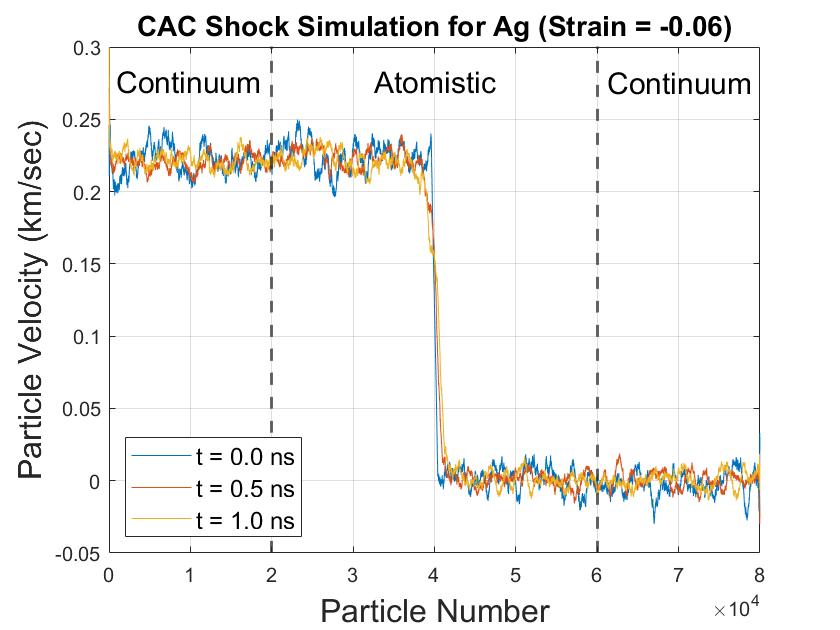}
                    \caption{}
                \end{subfigure}
                \begin{subfigure}{0.48\textwidth}
                    \includegraphics[width=\textwidth]{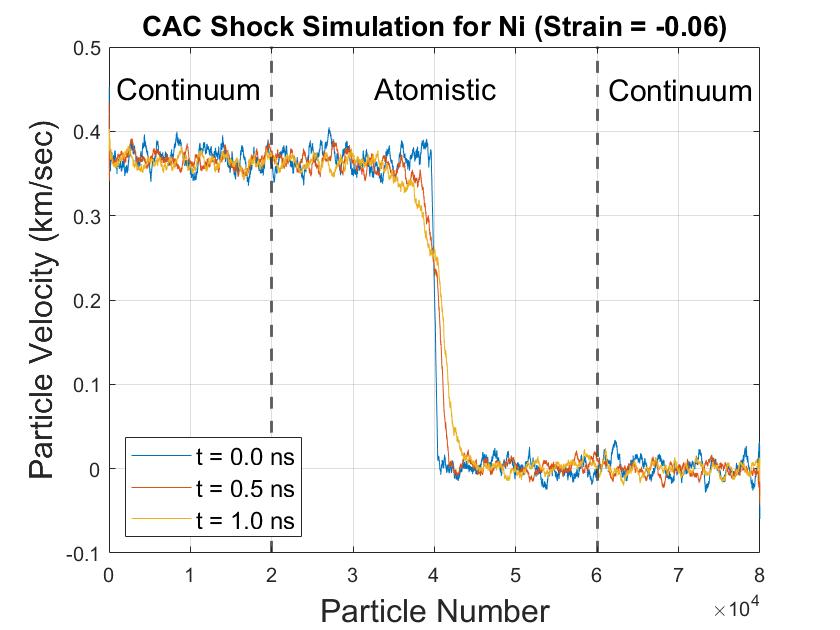}
                    \caption{}
                \end{subfigure}
                \caption{Velocity profiles of shock wave simulations performed with the conveyor technique for $\epsilon^+$ = -0.06. 
                Results are shown for the following four materials: (a) Cu, (b) Al, (c) Ag, and (d) Ni.}
                \label{Fig:CAC_ShockPlots_Conveyor}
            \end{figure}

            We first simulate propagating shock waves in the CAC domain using the conveyor moving window technique described in Sec. \ref{Sec: Conveyor technique}.
            For these simulations, the left and right coarse-scaled regions each contain 20,000 nodes with an element length of $6r_0$ for a total of 80,000 particles in the domain.
            Additionally, each atomistic TR band contains $100$ atoms where $\zeta = \zeta_0$ increases linearly throughout each band.
            The atomistic TRs are much longer than the force range to ensure that the WR achieves the correct temperature.
            Simulations are performed with Cu, Al, Ag, and Ni for final strains ($\epsilon^+$) ranging from -0.01 to -0.1, and each simulation is allowed to run for $1.0$ ns. 
            Results from four of these simulations for $\epsilon^+ = -0.06$ are shown in Fig. \ref{Fig:CAC_ShockPlots_Conveyor}.
            Here, the dotted lines represent the A-C interfaces, and we plot the velocity profile of the shock at $0.0$ ns, $0.5$ ns, and $1.0$ ns.
        
            For each of these shock simulations, $v^+$, $U_S$, and $\theta^+$ are obtained using the 3rd-order Eulerian thermoelastic equations from Sec. \ref{Sec: Shock equations} (Eqs. \ref{Eq: ClaytonEulerianParticleVelocity}, \ref{Eq: ClaytonEulerianShockVelocity}, and \ref{Eq: ClaytonEulerianTemperature} respectively). 
            The Gr\"{u}neisen constant $\Gamma_1$ and density $\rho_0$ are given in Sec. \ref{Sec: Interatomic potentials}, and $\hat{\Gamma}_{11}$ is determined from Eq. (\ref{Eq: ClaytonEulerianGruneisenConstant}).
            Additionally, the 2nd and 3rd order elastic constants are calculated from the equations developed by Born \cite{born1940stability}.
            These equations allow one to compute the elastic constants of cubic monatomic crystals whose atoms interact according to a central pairwise force law \cite{girifalco1959application,lincoln1967morse,singh2020third}:
            \begin{equation} \label{C11Equation}
                C_{11} = \frac{a^4}{2V} \sum_j m_j^4 D_j^2 \Pi(r_j)
            \end{equation}
            \begin{equation} \label{C111Equation}
                C_{111} = \frac{a^6}{2V} \sum_j m_j^6 D_j^3 \Pi(r_j).
            \end{equation}
            Here, \textit{a} is one-half the lattice parameter of the material, \textit{V} is the volume per atom ($V = 2a^3$ for FCC crystals), and $\Pi$ is the interatomic potential.
            Also, 
            \begin{equation}
                r_j = \left[m_j^2 + n_j^2 + l_j^2 \right]^{\frac{1}{2}}a
            \end{equation}
            where $m_j$, $n_j$, and $l_j$ are position coordinates of any atom in the lattice. 
            Finally, $D_j$ is an operator given as follows:
            \begin{equation}
                D_j = \frac{1}{r_j} \frac{d}{dr_j}
            \end{equation}
            
\begin{figure}[htpb]
                \centering
                \begin{subfigure}{0.48\textwidth}
                    \includegraphics[width=\textwidth]{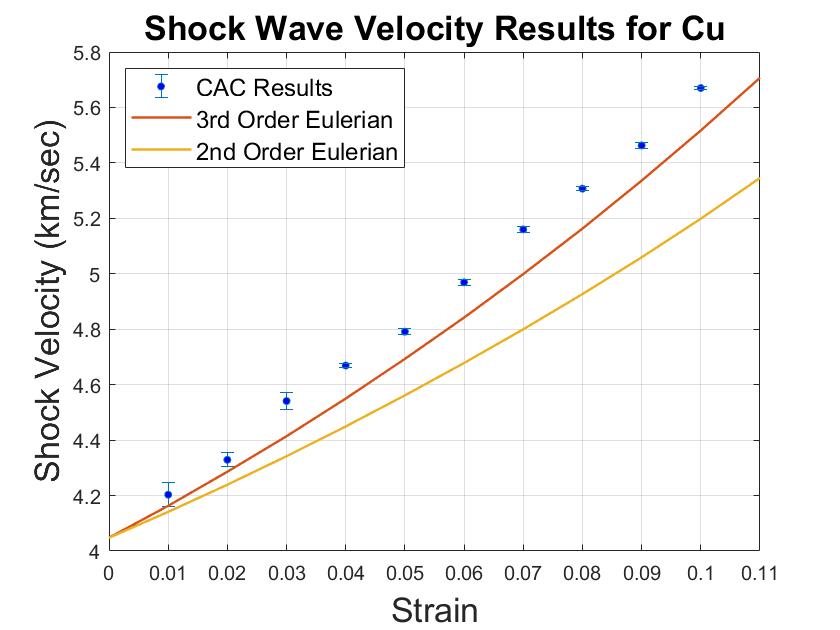}
                    \caption{}
                \end{subfigure}
                \begin{subfigure}{0.48\textwidth}
                    \includegraphics[width=\textwidth]{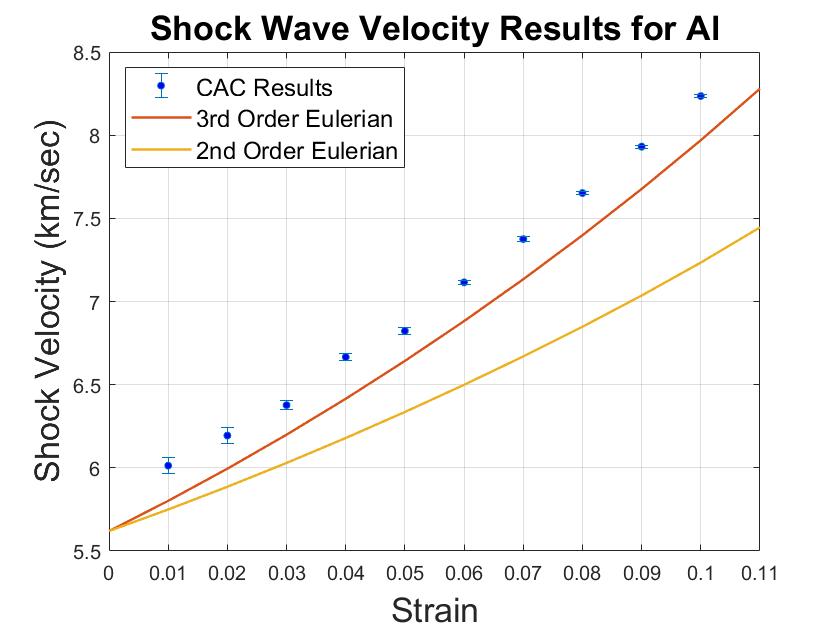}
                    \caption{}
                \end{subfigure}
                \\
                \begin{subfigure}{0.48\textwidth}
                    \includegraphics[width=\textwidth]{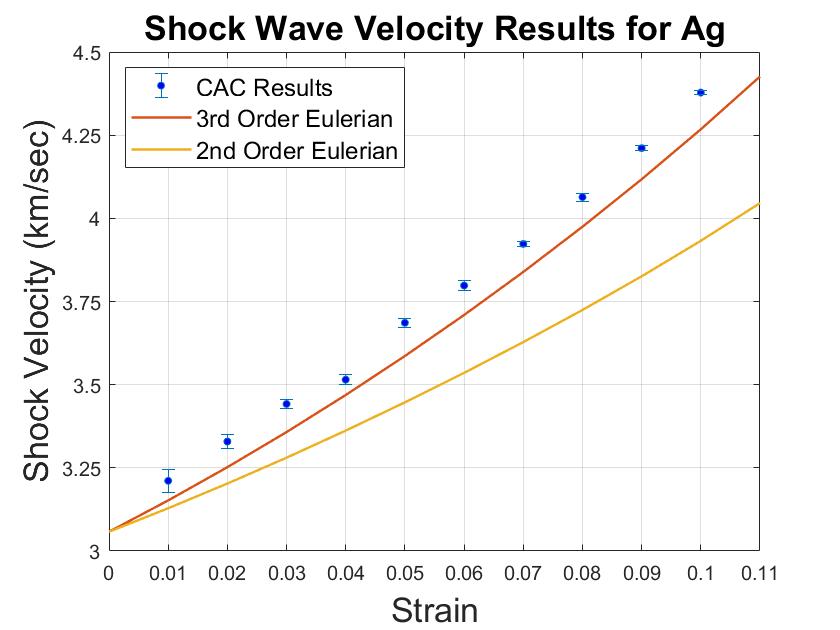}
                    \caption{}
                \end{subfigure}
                \begin{subfigure}{0.48\textwidth}
                    \includegraphics[width=\textwidth]{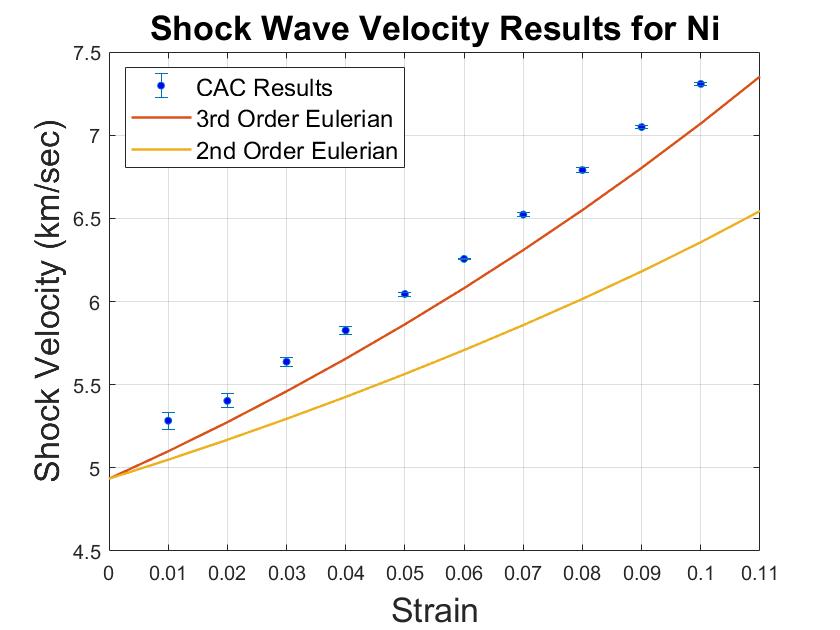}
                    \caption{}
                \end{subfigure}
                \caption{Simulated shock velocity ($U_S$) vs. strain ($\epsilon^+$) data for (a) Cu, (b) Al, (c) Ag, and (d) Ni.
                Analytical results from 2nd and 3rd order Eulerian theory are also shown.} 
                \label{Fig:CAC_ShockVelocityData_Conveyor}
            \end{figure}

            With the modified Morse potential, $r_0$ is defined as the equilibrium spacing along the close packed direction of the lattice.
            Therefore, we calculate $C_{11}$ and $C_{111}$ using Eqs. (\ref{C11Equation}) and (\ref{C111Equation}) by summing over the first nearest neighbors of the 1D chain, and we find $\hat{C}_{111}$ from these constants using Eq. (\ref{Eq: ClaytonEulerianC111}).
            Knowing $\Gamma_1$, $\hat{\Gamma}_{11}$, $\rho_0$, $C_{11}$, and $\Hat{C}_{111}$, we can then obtain the analytical shock velocity $U_S$ (as well as $v^+$ and $\theta^+$) required to begin the simulation.
            As seen in Fig. \ref{Fig:CAC_ShockPlots_Conveyor}, the shock front drifts over time from the center of the WR, and hence the simulated shock velocity deviates slightly from the analytical value.
            We fit the shock profile to a hyperbolic tangent function using MATLAB's Curve Fitting tool, and track the shock drift over 1.0 ns (with an equilibration time of 300 ps) to obtain the actual shock velocity.
            These results for all four elements over the full range of compressive strains studied can be seen in Fig. \ref{Fig:CAC_ShockVelocityData_Conveyor}.
            
            We observe in Fig. \ref{Fig:CAC_ShockVelocityData_Conveyor} that the shock velocity data obtained from the CAC simulations correspond well to the results obtained from 3rd-order Eulerian theory over the full range of strains.
            For all four materials, the actual shock velocity exceeds the analytical value by 2-3.5\%.
            Specifically, the average errors for Cu, Al, Ag, and Ni are given respectively as follows: 2.33\%, 3.33\%, 2.25\%, 3.23\%. 
            These errors are attributed to the analytical approximation of the elastic constants in Eqs. (\ref{C11Equation}) and (\ref{C111Equation}) (in fact, \cite{lincoln1967morse} found that these equations slightly under-predicted the experimental elastic constants of a bulk lattice at room temperature).
            The stiffer CAC results may also be due to thermal effects and including fourth order terms in the Eulerian equations could further reduce the differences between simulation and theory.
            Nonetheless, the drift in the shock front is nearly imperceptible for many hundreds of picoseconds which shows that such errors are fairly small for this type of problem.
            Therefore, the conveyor method maintains a stationary wave front at the center of the WR with minimal deviation from the theoretical shock velocity.
            
\begin{figure}[htpb]
                \centering
                \begin{subfigure}{0.48\textwidth}
                    \includegraphics[width=\textwidth]{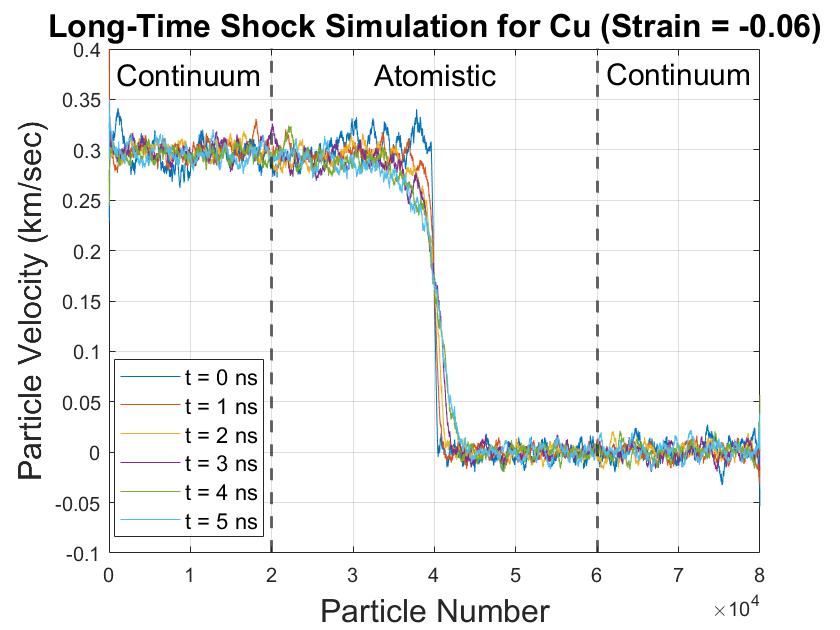}
                    \caption{}
                \end{subfigure}
                \begin{subfigure}{0.48\textwidth}
                    \includegraphics[width=\textwidth]{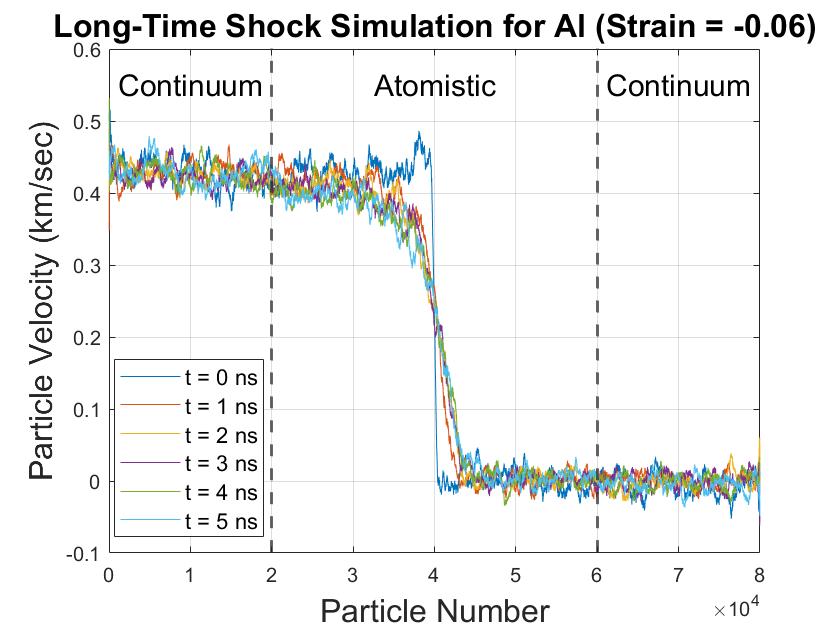}
                    \caption{}
                \end{subfigure}
                \\
                \begin{subfigure}{0.48\textwidth}
                    \includegraphics[width=\textwidth]{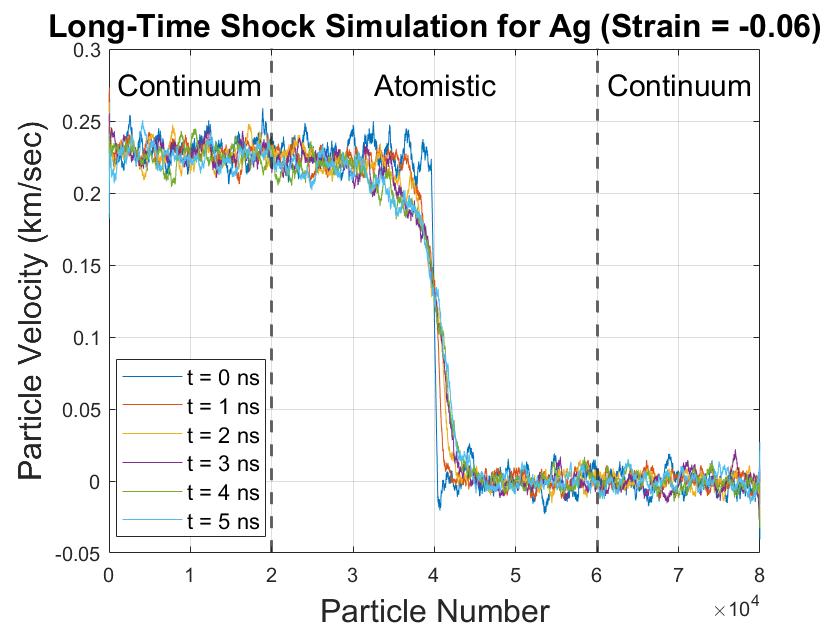}
                    \caption{}
                \end{subfigure}
                \begin{subfigure}{0.48\textwidth}
                    \includegraphics[width=\textwidth]{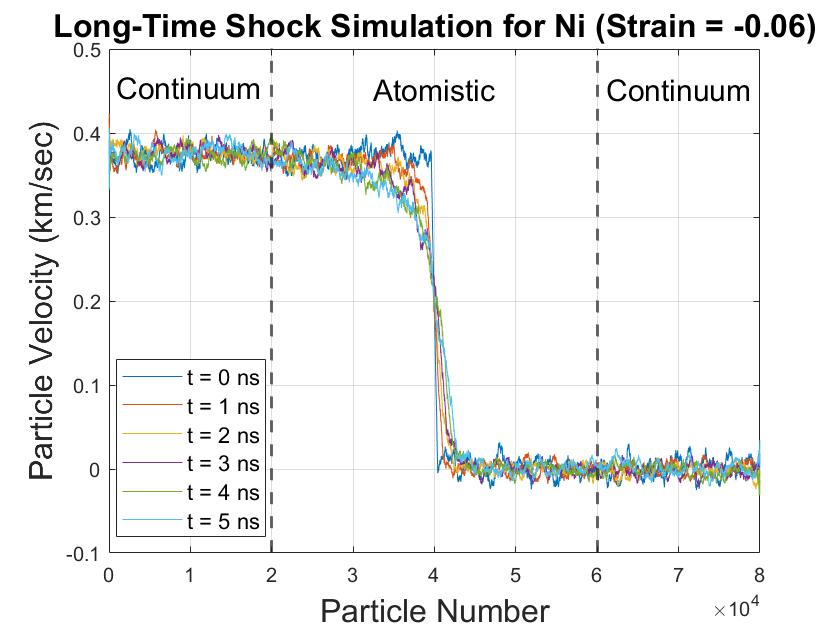}
                    \caption{}
                \end{subfigure}
                \caption{Velocity profiles of shock simulations performed with (a) Cu, (b) Al, (c) Ag, and (d) Ni for 5 ns ($\epsilon^+ = -0.06$).}
                \label{Fig:ShockPlot_LongTime}
            \end{figure}

            To confirm the ability of the CAC framework to track the propagating shock for very long simulation times with this moving window technique, we conduct simulations using Cu, Al, Ag, and Ni with $\epsilon^+ = -0.06$ for 5 ns, where $U_S$ and $v^+$ are assigned their computational values.
            Results from these simulations can be seen in Fig. \ref{Fig:ShockPlot_LongTime}, and in each case, we observe the shock front remain stationary at the center of the WR.
            Hence, the conveyor technique allows the CAC domain to track a propagating shock much longer than traditional NEMD simulations.
            Additionally, each of these domains contains 80,000 particles and covers a total length of nearly $70~\upmu\mathrm{m}$. 
            An equally-sized atomistic domain would be composed of 280,000 atoms and thus require $\sim$3.5x more memory.
            Therefore, the CAC framework with the conveyor technique significantly reduces the computational overhead of large-domain simulations.
        
        \subsection{Coarsen-refine method}
\begin{figure}[htpb]
                \centering
                \begin{subfigure}{0.48\textwidth}
                    \includegraphics[width=\textwidth]{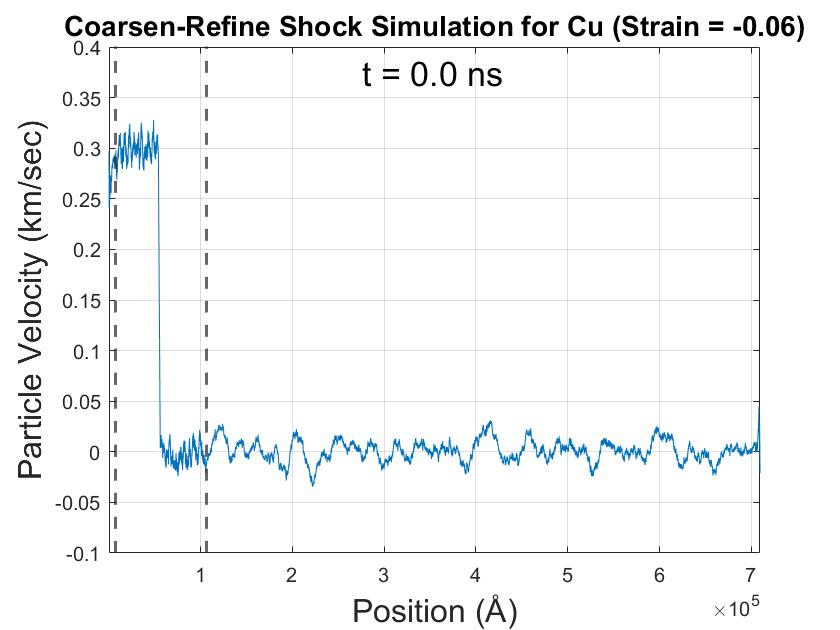}
                    \caption{}
                \end{subfigure}
                \begin{subfigure}{0.48\textwidth}
                    \includegraphics[width=\textwidth]{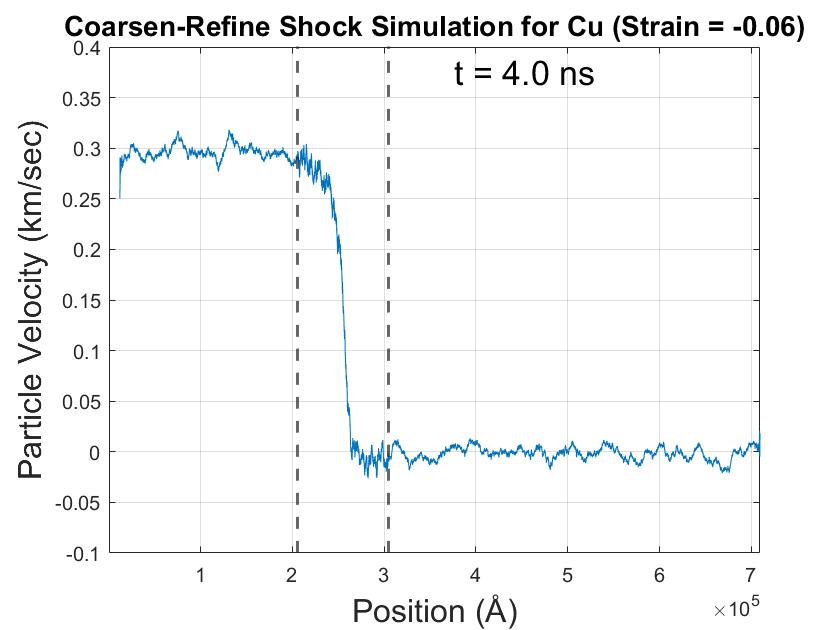}
                    \caption{}
                \end{subfigure}
                \\
                \begin{subfigure}{0.48\textwidth}
                    \includegraphics[width=\textwidth]{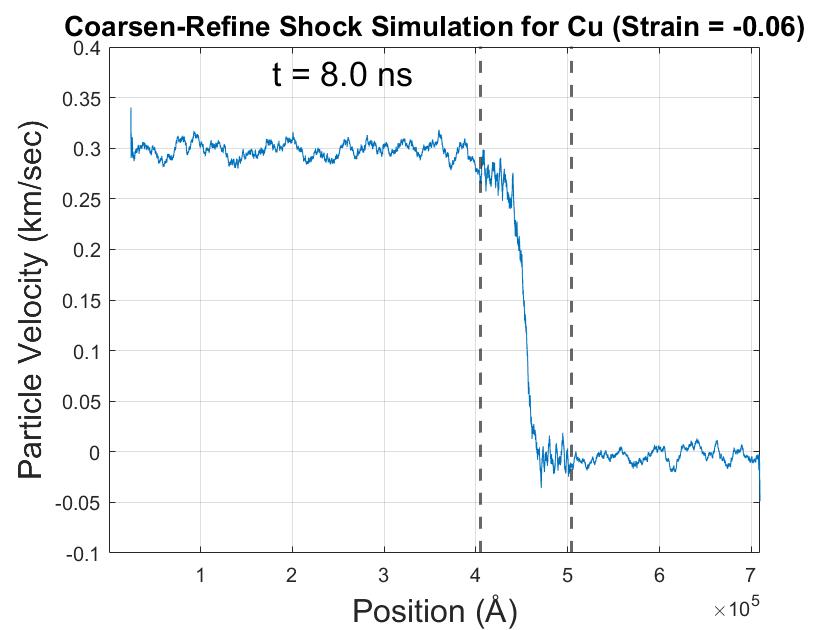}
                    \caption{}
                \end{subfigure}
                \begin{subfigure}{0.48\textwidth}
                    \includegraphics[width=\textwidth]{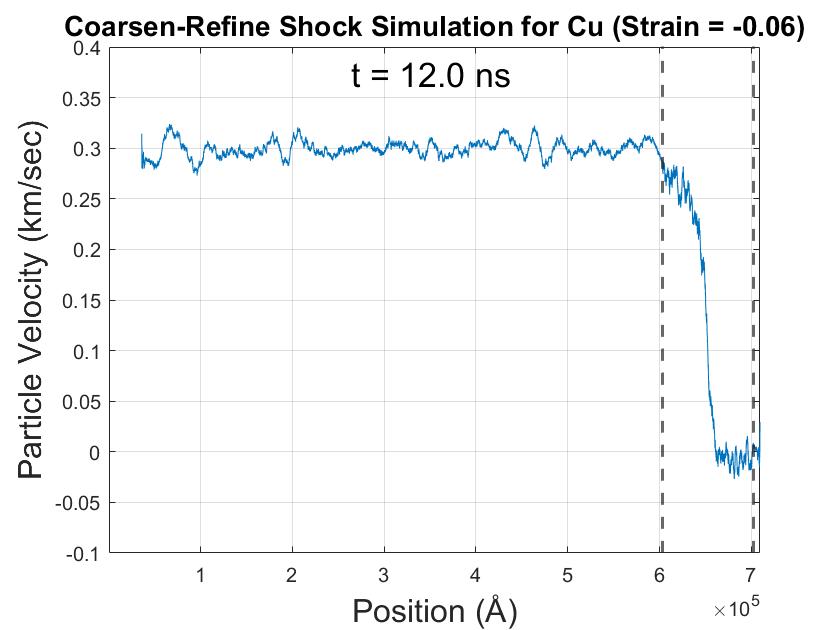}
                    \caption{}
                \end{subfigure}
                \caption{Velocity profile snapshots of a single propagating shock at the following times: (a) $0.0$ ns, (b) $4.0$ ns, (c) $8.0$ ns, and (d) $12.0$ ns.
                Here, $\epsilon^+$ = -0.06, and the atomistic domain follows the shock front using the coarsen-refine technique.
                The dotted lines represent the A-C interfaces.} 
                \label{Fig:CAC_ShockPlots_CoarsenRefine}
            \end{figure}

            Next, we track a propagating shock in the CAC domain using the coarsen-refine moving window technique described in Sec. \ref{Sec: Coarsen-refine technique}.
            In this case, however, the fine-scaled portion of the chain begins at the left boundary and is allowed to travel with the shock front through the entire domain by the simultaneous coarsening of the left continuum region and refinement of the right continuum region.
            Therefore, the shocked coarse-scaled region begins with 500 nodes, the unshocked coarse-scaled region begins with 39,500 nodes, and the fine-scaled region again contains 40,000 atoms for a total of 80,000 particles.
            The element length in each continuum region is $6r_0$ giving a total domain size of $\sim$ 709,367 \AA.
            As before, each atomistic TR band contains 100 atoms.
            During the simulation, the left continuum region will grow, the right continuum region will shrink, and the atomistic region will remain at a constant length.
            Results of a simulation performed with Cu for $\epsilon^+ = -0.06$ are shown in Fig. \ref{Fig:CAC_ShockPlots_CoarsenRefine}.
            
            In these plots, we observe the evolution of the shock wave over $12.0$ ns.
            The fine-scaled region successfully tracks the moving wave front through the entire CAC domain with minimal spurious behavior at the A-C interfaces.
            As expected, we observe compression of the material to the left of the shock wave while the rightmost point remains stationary.
            Hence, using the coarsen-refine moving window technique in a CAC framework allows us to model shock propagation over engineering length scales and time frames.
            We note that the results in Fig. \ref{Fig:CAC_ShockPlots_CoarsenRefine} can be directly compared to the results in Fig. \ref{Fig:ShockPlot_LongTime}a as both simulations follow a shock through Cu with a compressive strain of -0.06.
            In contrast to the conveyor technique, however, the shock can now travel through the entire domain because the lengths of the continuum regions change. 
            Previous studies have incorporated mesh refinement schemes into both finite element and atomistic-continuum frameworks \cite{berger1989local,xu2016mesh,tembhekar2017automatic,amor2021adaptive}.
            However, employing \textit{simultaneous} coarsening/refinement methods to model highly non-equilibrium events like  propagating shock waves is an active area of research.
            This work serves as a proof of concept for using such a technique in a multiscale setting.
            
        \subsection{Shock front structure}
\begin{figure}[htpb]
                \centering
                \includegraphics[width=0.55\textwidth]{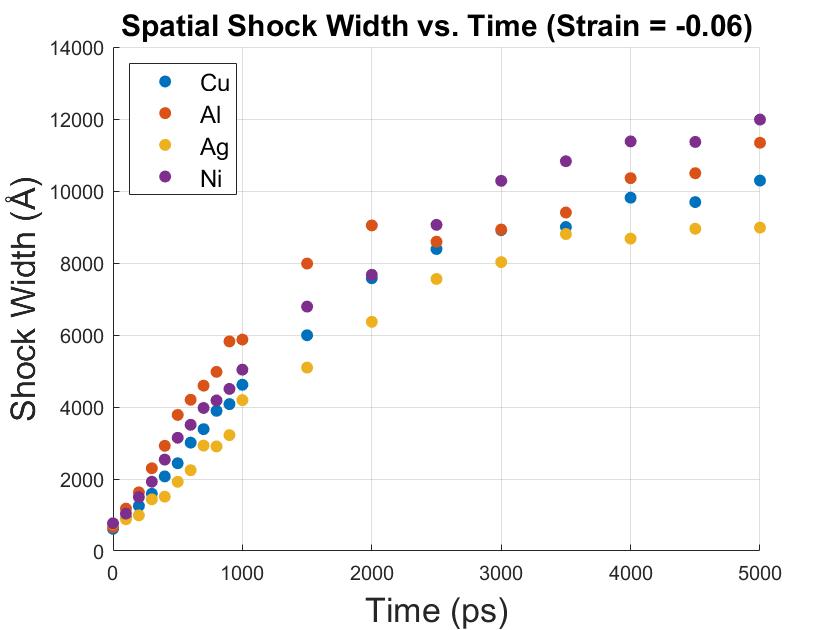}
                \caption{Growth in the spatial width of the shock wave front over 5 ns. Here, $\epsilon^+$ = -0.06, and results are shown for Cu, Al, Ag, and Ni.}
                \label{Fig:SpatialShockWidthVsTime}
            \end{figure}

            \added[id=R2,comment={1}]{In Figs. \ref{Fig:ShockPlot_LongTime} and \ref{Fig:CAC_ShockPlots_CoarsenRefine}, we observe that the spatial width of the shock front increases over time -- implying that the shock wave is unsteady.
            To understand this phenomenon further, we analyze the growth in shock front thickness over 5 ns for the four different materials.
            A plot of this data obtained for $\epsilon^+$ = -0.06 is shown in Fig. \ref{Fig:SpatialShockWidthVsTime}.
            These results are in qualitative agreement with results from previous NEMD studies which modeled weak shock waves through materials in a one-dimensional setting and also observed unsteady wave behavior \cite{holian1979molecular,holian1995atomistic,davis2020one,holian1998plasticity}.
            Additionally, we note that the growth rate of the shock width is approximately constant up to $\sim$1,000 ps and then slows down at higher time steps.
            This characteristic mirrors the results obtained previously in the moving window atomistic framework \cite{davis2020one}.
            Finally, from this data, we obtain a lower bound of $\sim$0.003 ns and an upper bound of $\sim$0.3 ns for the shock rise time across each material. 
            These values are very similar to the ranges observed in previous studies for small-domain FCC metals \cite{gahagan2000measurement,chhabildas1979rise}.}
            
        \subsection{Computational efficiency}
            Finally, we compare the computational cost of a shock simulation performed with CAC to an equally-sized MD simulation of a shock wave.
            MD simulations were conducted using an entirely fine-scaled domain.
            The runtime is calculated for a range of domain sizes, and the CAC framework always contains 40,000 atoms in the fine-scaled region.
            The results are summarized in Table \ref{Table: ComputationalEfficiency}.
            Here, the units represent hours of runtime required for every nanosecond of simulation time ($h/ns$), and each simulation was conducted with 40 processors.
            Notably, even the smallest CAC domain is already a factor of 2 faster than the pure MD system.
            For the longest domain, the speedup factor grows to nearly 6 even though there are only 4.67x as many atoms in the MD simulation as particles in the CAC simulation.
            This study illustrates benefits of using CAC with the moving window techniques to significantly reduce the computational time of large-scale, nonlinear simulations.
            \begin{table}[h]
                \centering
                \caption{Simulation costs of the CAC method against pure MD for various domain sizes. 
                The number of particles required in CAC and number of atoms required in MD for the given domain length is shown (the CAC system always contains 40,000 atoms in the fine-scaled region).}
                \begin{tabular}{||c  c  c  c  c  c||}
                \hline
                \textit{Domain size (\AA)} & Particles (CAC) & Atoms (MD) & \textit{h/ns (CAC)} & \textit{h/ns (MD)} & \textit{Speedup} \\
                \hline
                98,827 & 40,000 & 40,000 & 0.371 & 0.371 & 1.000 \\
                247,069 & 50,000 & 100,000 & 0.505 & 0.996 & 1.972 \\
                395,310 & 60,000 & 160,000 & 0.600 & 1.544 & 2.573 \\
                543,551 & 70,000 & 220,000 & 0.724 & 1.976 & 2.729 \\
                691,792 & 80,000 & 280,000 & 0.765 & 2.542 & 3.323 \\
                840,034 & 90,000 & 340,000 & 0.912 & 3.171 & 3.477 \\
                988,275 & 100,000 & 400,000 & 0.996 & 4.440 & 4.458 \\
                1,729,481 & 150,000 & 700,000 & 1.583 & 9.343 & 5.902 \\
                \hline 
                \end{tabular}
                \label{Table: ComputationalEfficiency}
            \end{table}
    
    \section{Conclusion} \label{Sec: Conclusion}
        In this paper, we developed a moving window multiscale framework using the CAC method to model shock wave propagation through a one-dimensional monatomic chain.
        Specifically, we studied the classic Riemann problem of a single propagating shock wave in an infinite medium. 
        We characterized the shock at the continuum level using the third-order nonlinear Eulerian thermoelastic equations for shock compression in anisotropic crystals \cite{clayton2013nonlinear,clayton2014shock}.
        We then incorporated two different moving window techniques into the framework which tracked the moving shock front in distinct ways.
        In the first method, the entire domain followed the wave front in a conveyor fashion, while in the second method, the fine-scaled region traveled through the chain by simultaneous coarsening and refinement.
        
        We performed many verification studies with the framework including the replication of phonon dispersion curves and the simulation of phonon wave packets.
        Shock wave studies revealed that the CAC framework could accurately model propagating shocks with velocities very similar to those predicted by third-order Eulerian continuum theory.
        Additionally, these studies showed that the conveyor technique could maintain the wave front in the middle of the fine-scaled region for very long simulation times.
        Finally, simulations performed with the coarsen-refine technique demonstrated that the fine-scaled region could travel with the shock through the domain and thus prevent the shock wave from encountering the atomistic-continuum interface.
        
        This work showcases the ability of the CAC framework to simulate highly non-equilibrium, transient events like shock waves through engineering-scale domains over realistic time scales.
        Atomistic methods have modeled many features of shock propagation in materials over the past several decades, but the scalability of such schemes is restricted by computational resources.
        While multiscale methods like CAC overcome size-scalability issues, studies which use these atomistic-continuum frameworks to model shock-like events have been limited.
        This is partly because fast-moving, nonlinear phenomena like shock waves can be hard to capture even in fairly large computational frameworks.
        When such events are simulated, the total runtime is nevertheless bounded because the shock will eventually leave the domain.
        The CAC method along with the two moving window techniques discussed in this paper significantly reduces the computational expense of such simulations and provides a means to model a propagating wave over extended time scales.
        The present formulation will also be very valuable when studying other phenomena in a 1D chain such as nonlinear transition waves, Hertzian contact, and phonon transport \cite{daraio2005strongly,kevrekidis2013interaction,deng2020nonlinear,chen2018passing,xiong2014prediction}.
        
        Future work will incorporate a wave passing technique into the CAC framework.
        One of the main issues hindering progress in the development of concurrent multiscale schemes is the reality of spurious wave reflections at the A-C interfaces.
        Previous work has utilized LD to allow short wavelength, high-frequency phonon waves to pass from the atomistic region to the continuum region \cite{chen2018passing}.
        We hope to expand upon this work and use such techniques to model complex phenomena like elastic waves emanating from a shock impact.
        Additionally, since transverse atomic motion and dislocations cannot be studied in a 1D setting, we intend to expand the current framework to higher dimensions and study events like phase transformations.
        In a 2D or 3D scheme, the WR would be either an area or volume and the A-C interfaces would be either lines or planes which could fluctuate in direction due to the moving window. 
        Additionally, the shock could travel at different angles when compared to the lattice orientation of the crystalline system.
        Despite these challenges, incorporating a coarsening/refinement technique into such a higher-dimensional framework would allow us to study effects at the shock front when it encounters phonon waves or other microstructural boundaries.

    \section{Acknowledgments} \label{Sec: Acknowledgments}
        This material is based upon work supported by the National Science Foundation under Grant No. $1950488$.
        Financial support was also provided by the U.S. Department of Defense through the National Defense Science and Engineering Graduate (NDSEG) Fellowship Program (F-$1656215698$). 
        Simulations were performed using the Easley computing cluster at Auburn University.
        The authors would like to thank Dr. John Clayton for valuable discussion during the development of the framework.

    \bibliographystyle{ieeetr}
    \bibliography{CAC1D_MW_Shocks}
    
    \appendix
    
    \section{Force vs. displacement tests} \label{App: Force vs. displacement tests}
\begin{figure}[htpb]
            \centering
            \begin{subfigure}{0.48\textwidth}
                \includegraphics[width=\textwidth]{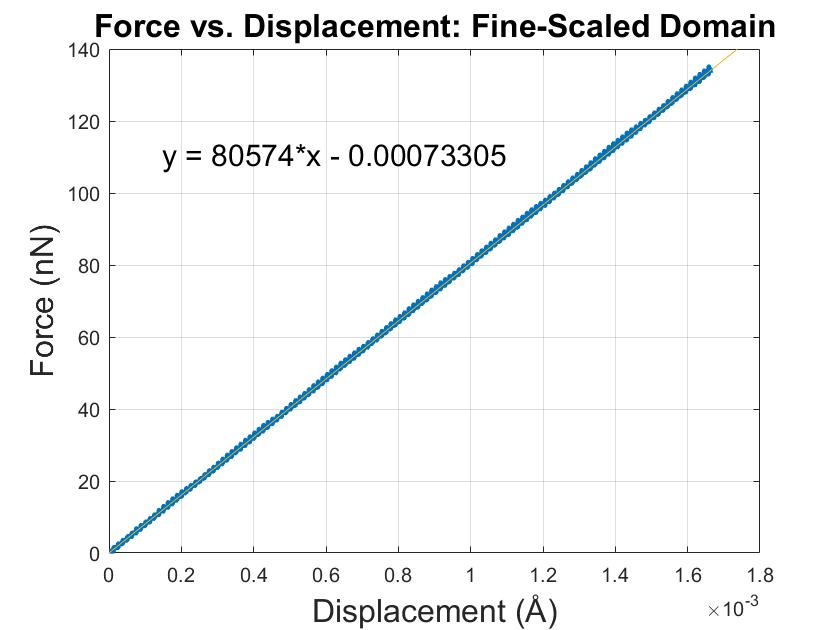}
                \caption{}
            \end{subfigure}
            \begin{subfigure}{0.48\textwidth}
                \includegraphics[width=\textwidth]{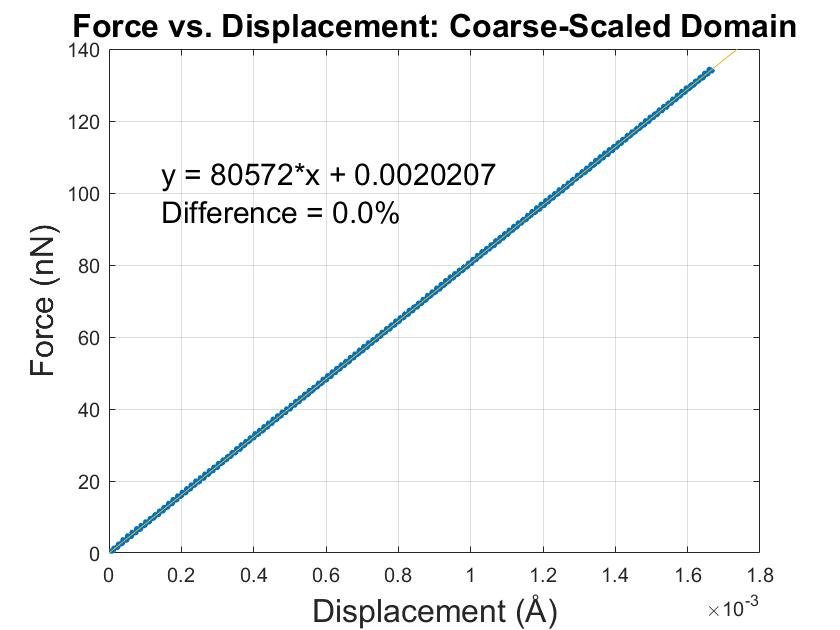}
                \caption{}
            \end{subfigure}
            \\
            \begin{subfigure}{0.48\textwidth}
                \includegraphics[width=\textwidth]{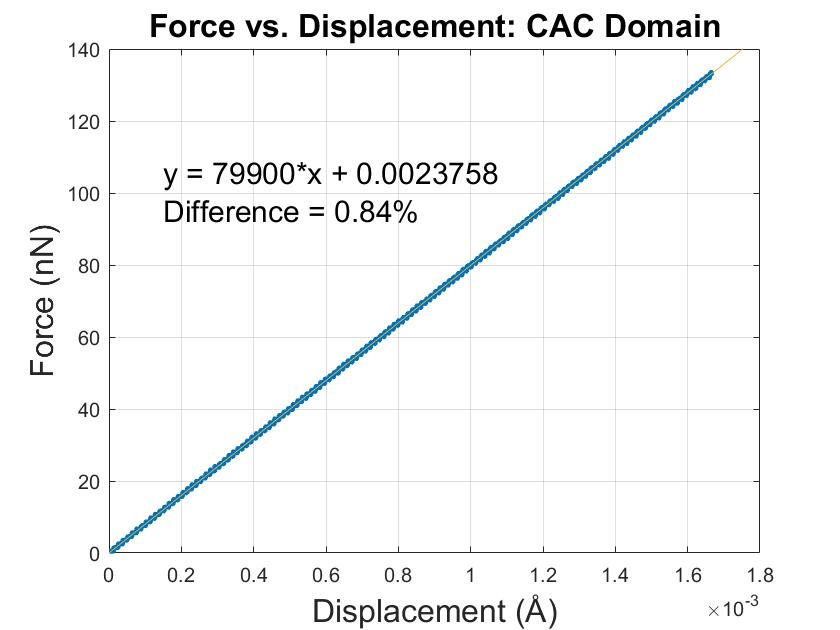}
                \caption{}
            \end{subfigure}
            \caption{Force vs. displacement test results for (a) fine-scaled, (b) coarse-scaled, and (c) CAC domains.}
            \label{Fig:FvDMorse}
        \end{figure}

        To ensure that the one-dimensional CAC framework achieves proper force matching at the A-C interfaces and thus does not produce any spurious wave phenomena, we perform force vs. displacement tests.
        This is accomplished by pulling the first particle in the chain very slowly at a constant rate and plotting the net force acting on it as a function of its absolute displacement over time.
        Specifically, we displace the particle at a rate of $1.6695$ $\times$ $10^{-7}$ \AA/ps for 10,000 ps.
        We perform these simulations using fine-scaled, coarse-scaled, and CAC domains with a total of 595 particles and lattice points in each system. 
        Hence, the fine-scaled domain consists of 595 atoms, the coarse-scaled domain consists of 100 nodes with each element having a length of $6r_0$, and the CAC domain consists of an inner atomistic region containing 360 atoms and two outer continuum regions each containing 20 nodes. 
        The results from these force vs. displacement studies can be seen in Fig. \ref{Fig:FvDMorse}.
        
        We observe that the coarse-scaled and CAC domains produce slopes which are nearly identical to the slope obtained from the fine-scaled domain with relative errors of 0.0\% and 0.84\% respectively.
        We note that even smaller relative errors were achieved with larger domain sizes.
        This implies that the spring constant is approximately the same in all three frameworks. 
        From these results, we conclude that the nodal integration and linear interpolation schemes used in our CAC framework yield accurate forces on the individual particles. 
        Therefore, Eq. (\ref{Eq: Internal Force Density}) is implemented correctly, and our system does not produce any spurious phenomena at the A-C boundaries.
        Additionally, the spring constant remains largely unchanged when transitioning from a fine-scaled domain to an equally-sized coarse-scaled or CAC domain.
    
    \section{Temperature equilibration tests} \label{App: Temperature equilibration}
        In this section, we verify that the CAC system from Fig. \ref{Fig:CACMWGeometry} can achieve the correct NVT ensemble in the undamped WR when the Langevin thermostat is applied to each TR.
        Specifically, we ensure that the temperature of the WR remains stable for various desired temperatures under equilibrium conditions.
        The input temperature $\theta_0$ is specified, and the particles initially have random velocities such that the CAC domain contains the correct total energy for an equilibrium system at $\theta_0$.
        Ideally, under equilibrium conditions, the system should achieve steady state with the appropriate equipartition of kinetic and potential energies.
        
\begin{figure}[htpb]
            \centering
            \includegraphics[width=0.55\textwidth]{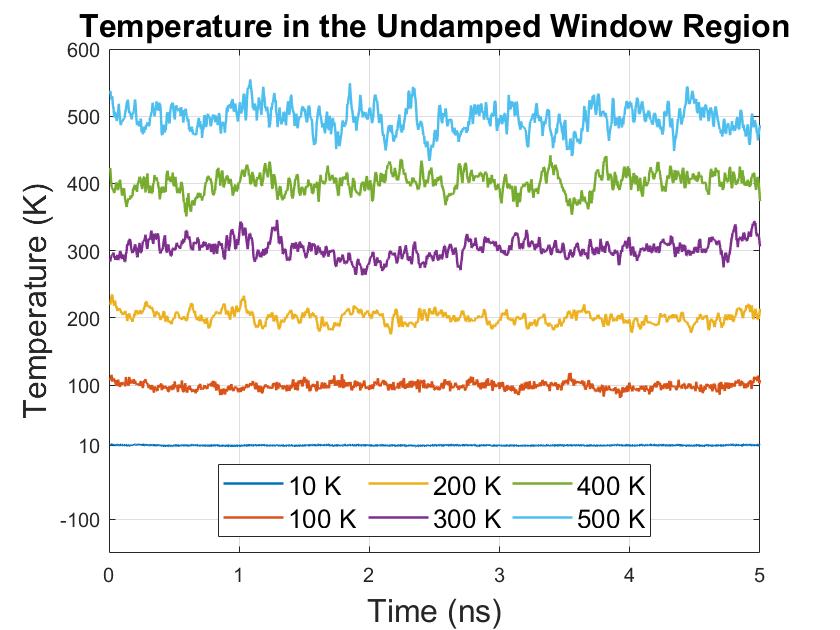}
            \caption{Temperature in the undamped WR vs. time.
            Here, the $Langevin$ thermostat is applied to the TRs for the following input temperatures: 10 K, 100 K, 200 K, 300 K, 400 K, and 500 K.}
            \label{Fig:CAC_Temperature_Equilibration}
        \end{figure}

        There are two main parameters associated with the CAC system from Fig. \ref{Fig:CACMWGeometry}: the length of each TR and the maximum damping parameter $\zeta_0$.
        To understand how these parameters affect the temperature equilibration of the WR, we performed studies over a range of different values.
        First, we found that each TR needed to be at least the range of the forces as a shorter length failed to dampen energetic pulses and resulted in wave reflections into the WR.
        Additionally, we found that the optimal value for $\zeta_0$ was one-half the Debye frequency ($\frac{1}{2} \omega_D$).
        A smaller $\zeta_0$ (weak damping) failed to achieve a canonical ensemble in the WR, while a larger $\zeta_0$ (hyper damping) resulted in large fluctuations in the WR \cite{holian1995thermostatted}.
        These results are consistent with results previously obtained for the CADD framework \cite{qu2005finite}.
        Therefore, we always use a maximum damping of $\frac{1}{2}\omega_D$ with the Langevin thermostat.
        Results for the temperature equilibration studies can be seen in Fig. \ref{Fig:CAC_Temperature_Equilibration}.
        
        For each of these simulations, the CAC domain contains a total of 1,000 particles with 100 nodes in each coarse-scaled TR, 50 atoms in each fine-scaled TR, and 700 atoms in the WR. 
        We then evolve the system for 5 ns and calculate the temperature of the undamped WR at each time step using the equipartition theorem. 
        The temperature in each TR remains steady at the input temperature $\theta_0$ and thus acts as a constant-temperature reservoir.
        In Fig. \ref{Fig:CAC_Temperature_Equilibration}, we plot the temperature in the WR vs. time for the following input values: $10$ K, $100$ K, $200$ K, $300$ K, $400$ K, and $500$ K.
        In each case, we observe the temperature stabilize after a short time and achieve a steady state around the respective input temperature. 
        We note that the variance in the temperature increases with larger input values because the particles oscillate more rapidly at higher temperatures.
        This effect can also be seen in Table \ref{Table: AverageTemperatureWindowRegion} which shows the time-average temperature and its corresponding standard deviation in the WR for each simulation.
        These results confirm that when an input temperature $\theta_0$ is applied to each TR, the undamped WR maintains an equilibrium steady state around this value.
        \begin{table}[h]
            \centering
            \caption{Time-average temperature in the WR and its associated standard deviation for various input temperatures $\theta_0$ applied to each TR.}
            \begin{tabular}{||c  c  c||}
            \hline
            $\theta_0$ (K) & $\langle \theta \rangle$ & $\langle \delta \theta \rangle$ \\
            \hline
            10 & 10.19 & 0.52 \\ 
            100 & 98.64 & 5.37 \\ 
            200 & 200.65 & 11.24 \\ 
            300 & 302.72 & 16.52 \\ 
            400 & 400.66 & 19.35 \\ 
            500 & 496.39 & 24.56 \\ 
            \hline 
            \end{tabular}
            \label{Table: AverageTemperatureWindowRegion}
        \end{table}

\end{document}